\documentclass[format=acmsmall, review=false]{acmart}

\usepackage{acme20} % For EC submission, comment this out and uncomment the next line 

\usepackage[utf8]{inputenc}
\usepackage{amsmath}
\usepackage{array}
\usepackage{graphicx}
\usepackage[english]{babel}
\usepackage[utf8]{inputenc}
\usepackage{algorithm}
\usepackage[noend]{algpseudocode}

\usepackage{enumitem}
\usepackage{wasysym}
\usepackage{framed}
\usepackage{pgfgantt,rotating}
\usepackage{subfloat}
\usepackage{multirow}
\usepackage{blindtext}

\usepackage[most]{tcolorbox}

\title{Morpheus Consensus: \newline Excelling on trails and autobahns}

\author{Andrew Lewis-Pye}
\affiliation{%
  \institution{\department{Department of Mathematics} \institution{London School of Economics} \city{London} \country{UK}}
}
\email{a.lewis7@lse.ac.uk}

\author{Ehud Shapiro}
\affiliation{%
 \institution{\department{Data Science Institute} \institution{London School of Economics} \city{London} \country{UK}} ~and
  \institution{\institution{Weizmann Institute of Science}
 \city{Rehovot} \country{Israel}}
}
\email{udi.shapiro@gmail.com}

\date{November 2024}

\begin{abstract} Recent research in consensus has often focussed on protocols for State-Machine-Replication (SMR) that can handle high throughputs. Such state-of-the-art protocols (generally DAG-based) induce undue overhead when the needed throughput is low, or else exhibit unnecessarily-poor latency and communication complexity during periods of low throughput.

Here we present Morpheus Consensus, which naturally morphs from a quiescent low-throughput leaderless blockchain protocol to a high-throughput leader-based DAG protocol and back, excelling in latency and complexity in both settings. During high-throughout, Morpheus pars with state-of-the-art DAG-based protocols, including Autobahn \cite{giridharan2024autobahn}.  During low-throughput, Morpheus exhibits competitive complexity and lower latency than standard protocols such as PBFT \cite{castro1999practical} and Tendermint \cite{buchman2016tendermint,buchman2018latest}, which in turn do not perform well during high-throughput. 

The key idea of Morpheus is that as long as blocks do not conflict (due to Byzantine behaviour, network delays, or high-throughput simultaneous production) it produces a forkless blockchain, promptly finalizing each block upon arrival.  It assigns a leader only if one is needed to resolve conflicts, in a manner and with performance not unlike Autobahn.
\end{abstract}

\begin{document}

\maketitle

\section{Introduction}  \label{intro} 
Significant investment in blockchain technology has recently led to renewed interest in research on consensus protocols. Much of this research is focussed on developing protocols that operate efficiently `at scale'. In concrete terms, this means looking to design protocols that can handle a high throughput (i.e.\ high rate of incoming transactions) with low latency (i.e.\ quick transaction finalization), even when the number of processes (validators) carrying out the protocol is large. 
%Recent protocols, such as Sailfish \cite{} and Autobahn \cite{}

\vspace{0.2cm} 
\noindent \textbf{Dealing efficiently with low and high throughput}. While blockchains may often need to handle high throughputs, it is not the case that \emph{all} blockchains need to deal with high throughput \emph{all} of the time. For example, various `subnets' or `subchains' may only have to deal with high throughputs infrequently, and should ideally be optimised to deal also with periods of low throughput.  The motivation for the present paper therefore stems from a real-world need for consensus protocols that deal efficiently with both high and low throughputs. Specifically, we are interested in a setting where: 
\begin{enumerate} 
\item The processes/validators may be few, but could be up to a few hundred in number. 
\item The protocol should be able to handle periods of asynchrony, i.e.\ should operate efficiently in the partially synchronous setting. 
\item The protocol is required to have optimal resilience against Byzantine adversaries, i.e.\ should be live and consistent so long as less than 1/3 of processes display Byzantine faults, but should be optimised to deal with the `normal case' that processes are not carrying out Byzantine attacks and that faults are benign (crash or omission failures). 
\item There are expected to be some periods of  high throughput, meaning that the protocol should ideally match the state-of-the-art during such periods. 
\item Often, however, throughput will be low. This means the protocol should also be optimised to give the lowest possible latency during periods of low throughput. 
\item Ideally, the protocol should be `leaderless' during periods of low throughput:  the use of leaders is to be avoided if possible, since, even without malicious action,  leaders who are offline/faulty may cause significant increases in latency.
\item Ideally, the protocol should also be `quiescent', i.e.\ there should be no need for the sending and storing of new messages when new transactions are not being produced. 
\item Transactions may come from \emph{clients} (not belonging to the list of processes/validators), but will generally be produced by the processes themselves.  
\end{enumerate}   

\vspace{0.2cm} 
\noindent \textbf{The main contribution of this paper}. We introduce and analyse the Morpheus protocol, which is designed for the setting described above. The protocol is quiescent and has the following properties during periods of low throughput: 
\begin{itemize} 
 \item It is leaderless, in the sense that transactions are finalized without the requirement for involvement by leaders.
 \item  Transactions are finalized in time $3\delta$, where $\delta$ is the actual (and unknown) bound on message delays after GST.\footnote{The partially synchronous setting and associated notions such as GST are formally defined in Section \ref{setup}.} As explained in Section \ref{metrics}, this more than halves the latency of existing DAG-based protocols and variants such as Autobahn \cite{giridharan2024autobahn} for the low throughput case, and even decreases latency by at least $\delta$ when compared with protocols such as PBFT (and even if we suppose leaders for those protocols are non-faulty), since the leaderless property of our protocol negates the need to send transactions to a leader before they can be included in a block.
 \item  A further advantage over protocols such as PBFT and Tendermint is that crash failures by leaders are not able to impact latency during periods of low throughput. 
 \end{itemize} 

During periods of high throughput, Morpheus is very similar to Autobahn, and so inherits the benefits of that protocol. In particular: 
\begin{itemize} 
 \item It has the same capability to deal with high throughput as DAG-based protocols and variants such as Autobahn, and has the same ability to recover quickly from periods of asynchrony  (`seamless recover' in the language of Autobahn). 
 \item It has the same latency as Autobahn during high throughput, matching the latency of Sailfish \cite{shrestha2024sailfish}, which is the most competitive existing DAG-based protocol in terms of latency. 
 \item As detailed in Section \ref{metrics},  Morpheus has the same advantages as Autobahn in terms of communication complexity when compared to DAG-based protocols such as Sailfish, DAG-Rider \cite{keidar2021all}, Cordial Miners \cite{keidar2022cordial}, Mysticeti \cite{babel2023mysticeti} or Shoal \cite{spiegelman2023shoal}.
% \item  Another advantage of Autobahn over the latter DAG-based protocols (and also over protocols like PBFT and Tendermint), also inherited by Morpheus, is its ability to deal with significant network partitions: during network partitions in which one half of the network does not receive messages from the other, DAG-based protocols such as those listed above may stall production of the DAG, since each new block for a given `layer' is required to point to $n-f$ blocks from the previous layer.\footnote{Here, $n$ is the number of processes, while $f$ is the given bound on the number of faulty processes}. With Autobahn and Morpheus, each process is able to continue producing blocks during any network partition. Upon resumption of synchrony, other processes need only receive their most recent blocks to continue with the task of consensus.
 \end{itemize}  

Of course, much of the complexity in designing a protocol that operates efficiently in both low and high throughput settings is to ensure a smooth transition and consistency between the different modes of operation that the two settings necessitate. 

\vspace{0.2cm}
\noindent \textbf{Further contributions of the paper}. In Section \ref{esmr}, we also formalise the task of \emph{Extractable SMR}, as an attempt to  make explicit certain implicit assumptions that are often made by papers in the area. While State-Machine-Replication (SMR) requires correct processes to finalize \emph{logs} (sequences of transactions) in such a way that \emph{consistency} and \emph{liveness} are satisfied,   it is well understood in the community that some papers describing protocols for SMR   specify protocols that \emph{do not} actually aim to explicitly ensure all correct processes receive all finalized blocks (required for liveness). Roughly, the protocol instructions suffice instead to ensure \emph{data availability}  (that each finalized block is received by at least one correct process), and then the protocol is required to establish a total ordering on transactions that can be extracted via further message exchange, given data availability. Liveness is therefore only achieved after further message exchange (and via some unspecified method), which (while a trivial addition if one does not consider communication complexity) is not generally taken into account when calculating message complexity. 

In Hotstuff \cite{yin2018hotstuff}, for example, one of the principal aims is to ensure linear message complexity within \emph{views}. Since this precludes all-to-all communication within views, a Byzantine leader may finalize a block of transactions in a given view without certain correct processes even receiving the block. Those correct processes must eventually receive the block for liveness to be satisfied, but the protocol instructions do not explicitly stipulate the mechanism by which this should be achieved. While ensuring that all correct processes receive the block is trivial if one is not concerned with communication complexity (e.g., just have each correct process broadcast each finalized block they observe, together with a quorum certificate verifying that the block is finalized), the messages required to do so are not counted when analyzing message complexity. The obvious methods of ensuring that the block is propagated to all correct processes will require more than linear communication complexity, which undermines the very point of the Hotstuff protocol. 

The question arises, ``what precisely is the task being achieved by such protocols if they do not satisfy liveness without further message exchange (and so actually fail to achieve the task of SMR with the communication complexity computed)''. We assert that the task of Extractable SMR is an appropriate formalisation of the task being achieved, and hope that the introduction of this notion is a contribution of independent interest. 

\vspace{0.2cm}
\noindent \textbf{The structure of the paper}. The paper structure is as follows: 
\begin{itemize} 
\item Section \ref{setup} describes the basic model and definitions.
\item  Section \ref{esmr} formalises the task of Extractable SMR. 
\item Section \ref{intuition} gives the intuition behind the Morpheus protocol. 
\item Section \ref{formalspec} gives the formal specification of the protocol. 
\item Section \ref{cl} formally establishes consistency and liveness. 
\item Section \ref{metrics} analyses communication complexity and latency, and makes comparisons with the state-of-the-art. 
\item Section \ref{rw} discusses related work. 
\end{itemize}

\section{The setup} \label{setup} 

We consider a set $\Pi= \{ p_0,\dots, p_{n-1} \}$ of $n$ processes. Each process $p_i$ is told $i$ as part of its input. We consider an adaptive adversary, which chooses  a set of at most $f$ processes to corrupt during the execution, where $f$ is the largest integer less than $n/3$. A process that is corrupted by the adversary  is referred to as \emph{Byzantine} and may behave arbitrarily, subject to our cryptographic assumptions (stated below).  Processes that are not Byzantine are \emph{correct}.

\vspace{0.2cm} 
\noindent \textbf{Cryptographic assumptions}. Our cryptographic assumptions are standard for papers in distributed computing. Processes communicate by point-to-point authenticated channels. We use a cryptographic signature scheme, a public key infrastructure (PKI) to validate signatures,  a threshold signature scheme \cite{boneh2001short,shoup2000practical}, and a cryptographic hash function $H$. The threshold signature scheme is used to create a compact signature of $m$-of-$n$ processes, as in other consensus protocols \cite{yin2019hotstuff}. In this paper, $m=n-f$ or $m=f+1$.  The size of a threshold signature is $O(\kappa)$, where $\kappa$ is a security parameter, and does not depend on $m$ or $n$.
We assume a computationally bounded adversary. Following a common standard in distributed computing and for simplicity of presentation (to avoid the analysis of negligible error probabilities), we assume these cryptographic schemes are perfect, i.e.\ we restrict attention to executions in which the adversary is unable to break these cryptographic schemes.  
 Hash values are thus assumed to be unique.

 \vspace{0.2cm} 
\noindent \textbf{Message delays}. We consider a discrete sequence of timeslots $t\in \mathbb{N}_{\geq 0}$ in the partially synchronous setting: for some known bound $\Delta$ and unknown \emph{Global Stabilization Time} (GST),  a message sent at time $t$ must arrive by time $\max\{\text{GST},t\} + \Delta$. The adversary chooses GST and also message delivery times, subject to the constraints already specified. We write $\delta$ to denote the actual (unknown) bound on message delays after GST, noting that $\delta$ may be significantly less than the known bound $\Delta$.

\vspace{0.2cm} 
\noindent \textbf{Clock synchronization}.
We do not suppose that the clocks of correct processes are synchronized. For the sake of simplicity, however, we do suppose that the clocks of correct processes all proceed in real time, i.e.\ if $t'>t$ then the local clock of correct $p$ at time $t'$ is $t'-t$ in advance of its value at time $t$. This assumption is made only for the sake of simplicity, and our arguments are easily adapted to deal with a setting in which there is a known upper bound on the difference between the clock speeds of correct processes after GST. We suppose all correct processes begin the protocol execution before GST.  A correct process may begin the protocol execution with its local clock set to any value. 

%\vspace{0.2cm} 
%\noindent \textbf{Notation concerning executions and received messages}. We use the following notation: 
%\begin{itemize} 
%\item $M^{\mathcal{E}}_{p}(t)$ denotes the set of messages received by process $p$ by timeslot $t$ in execution $\mathcal{E}$;
%\item  $M_c^{\mathcal{E}}(t)$  denotes the set of all messages received by any correct process by timeslot $t$ in execution $\mathcal{E}$, 
%\item $M^{\mathcal{E}}$  denotes the set of all messages received by any process in execution $\mathcal{E}$. 
%\item  We write  $\mathcal{E}_{\leq t}$ to denote the execution $\mathcal{E}$ up to timeslot $t$.
%\end{itemize} 
 
% \vspace{0.2cm} 
%\noindent \textbf{Notation concerning executions and received messages}. We write $M^{\mathcal{E}}_{p}(t)$ to denote the set of messages received by process $p$ by timeslot $t$ in execution $\mathcal{E}$, $M_c^{\mathcal{E}}(t)$ to denote the set of all messages received by any correct process by timeslot $t$ in execution $\mathcal{E}$, and
% $M^{\mathcal{E}}$ to denote the set of all messages received by any process in execution $\mathcal{E}$.  We write  $\mathcal{E}_{\leq t}$ to denote the execution $\mathcal{E}$ up to timeslot $t$.

\vspace{0.2cm} 
\noindent \textbf{Transactions}. Transactions are messages of a distinguished form. For the sake of simplicity, we consider a setup in which each process produces their own transactions, but one could also adapt the presentation to a setup in which transactions are produced by clients who may pass transactions to multiple processes. 
%We allow that processes may also send signed transactions to each other.  

\section{Extractable SMR} \label{esmr} 

\noindent \textbf{Informal discussion}. State-Machine-Replication (SMR) requires correct processes to finalize \emph{logs} (sequences of transactions) in such a way that \emph{consistency} and \emph{liveness} are satisfied. As noted in Section \ref{intro}, however, for many papers describing protocols for SMR, the  explicit instructions of the protocol \emph{do not} actually suffice to ensure liveness without further message exchange, potentially impacting calculations of message complexity and other measures. Roughly, the protocol instructions do not explicitly ensure that all correct processes receive all finalized blocks, but rather ensure \emph{data availability} (that each finalized block is received by at least one correct process), and then the protocol is required to establish a total ordering on transactions that can be extracted via further message exchange, given data availability.  Although it is clear that the protocol can be used to solve SMR given some (as yet unspecified) mechanism for message exchange, the protocol itself does not solve SMR. So, what exactly is the task that the protocol solves?

\vspace{0.2cm} 
\noindent \textbf{Extractable SMR (formal definition)}. If $\sigma$ and $\tau$ are strings, we write $\sigma \subseteq \tau$ to denote that $\sigma$ is a prefix  of $\tau$. We say $\sigma$ and $\tau$ are \emph{compatible} if $\sigma\subseteq \tau$ or $\tau\subseteq \sigma$. If two strings are not compatible, they are \emph{incompatible}.   If $\sigma$ is a sequence of transactions, we write $\mathtt{tr}\in \sigma$ to denote that the transaction $\mathtt{tr}$ belongs to the sequence $\sigma$. 

   If $\mathcal{P}$ is a protocol for \emph{extractable SMR}, then it must specify a function $\mathcal{F}$ that maps any set of messages to a sequence of transactions. Let $M^*$ be the set of all messages that are received by at least one (potentially Byzantine) process during the execution. For any timeslot $t$, let $M(t)$ be the set of all messages that are received by at least one correct process at a timeslot $\leq t$. We require the following conditions to hold:

\vspace{0.1cm} 
\noindent \emph{Consistency}.  For any $M_1$ and $M_2$, if $M_1\subseteq M_2 \subseteq M^*$, then $\mathcal{F}(M_1)\subseteq \mathcal{F}(M_2)$.

%\noindent (i) (No divergence.) $\mathcal{F}(M_1)$ and $\mathcal{F}(M_2)$ are compatible.  

%\vspace{0.1cm} 
%\noindent \emph{Client verifiability}.  If $p$ is correct then, for every $t$, there must exist  $M\subseteq M^{\mathcal{E}}_{p}(t)$ which is a \emph{certificate} for $\mathtt{log}^{\mathcal{E}}_p(t)$: $M$ is such a certificate if,  for every $\mathcal{E}'$ consistent with the setting and every correct $p'$ and $t'$ such that  $M \subseteq M^{\mathcal{E}'}_{p'}(t')$, $\mathtt{log}^{\mathcal{E}'}_{p'}(t')\supseteq \mathtt{log}^{\mathcal{E}}_p(t)$. (See the discussion below for further motivation.)

\vspace{0.1cm} 
\noindent \emph{Liveness}. If correct $p$ produces the transaction $\mathtt{tr}$, there must exist $t$ such that $\mathtt{tr}\in \mathcal{F}(M(t))$. 
% If this condition holds for $\ell$ and a specific $t$, then we also say that \emph{liveness holds with parameter $\ell$ at $t$}.  

\vspace{0.2cm} Note that consistency suffices to ensure that, for arbitrary $M_1,M_2\subseteq M^*$, $\mathcal{F}(M_1)$ and $\mathcal{F}(M_2)$ are compatible. To see this, note that, by consistency, $ \mathcal{F}(M_1) \subseteq \mathcal{F}(M_1 \cup M_2)$ and $\mathcal{F}(M_2)\subseteq \mathcal{F}(M_1\cup M_2)$. 

\vspace{0.2cm} 
\noindent \textbf{Converting protocols for Extractable SMR to protocols for SMR}. In this paper, we focus on the task of Extractable SMR. One way to convert a protocol for Extractable SMR into a protocol for SMR is to assume the existence of a gossip network, in which each process has some (appropriately chosen) constant number of neighbors. Using standard results from graph theory (\cite{bollobas2013modern} Chapter 7), one can assume correct processes form a connected component: this assumption requires classifying some small number of disconnected processes that would otherwise be correct as Byzantine. If each correct process gossips each `relevant' protocol message, then all such messages will eventually be received by all correct processes. Overall, this induces an extra communication cost per message which is only linear in $n$. 
Of course, other approaches are also possible, and in this paper we will remain agnostic as to the precise process by which SMR is achieved from Extractable SMR. 

\section{Morpheus: The intuition} \label{intuition} 

In this section, we informally describe the intuition behind the protocol. The protocol may be described as `DAG-based', in the sense that each block may \emph{point to} more than one previous block via the use of hash pointers. The blocks \emph{observed} by any block $b$ are $b$ and all those blocks observed by blocks that $b$ points to. The set of blocks observed by $b$ is denoted by $[b]$. If neither of $b$ and $b'$ observe each other, then these two blocks are said to \emph{conflict}. Blocks will be of three kinds: there exists a unique \emph{genesis} block $b_g$ (which observes only itself), and all other blocks are either \emph{transaction} blocks or \emph{leader} blocks. 

\vspace{0.2cm} 
\noindent \textbf{The operation during low throughput}. Roughly, by the `low throughput mode', we mean a setting in which processes produce blocks of transactions infrequently enough that correct processes agree on the order in which they are received, meaning that transaction blocks can be finalized individually upon arrival. Our aim is to describe a protocol that finalizes transaction blocks with  low latency in this setting, and without the use of a leader: the use of leaders is to be avoided if possible, since leaders who are offline/faulty may cause significant increases in latency. The way in which Morpheus operates in this setting is simple: 
\begin{enumerate} 
\item Upon having a new transaction block $b$ to issue, a process $p_i$ will send $b$ to all  processes. 
\item If they have not seen any blocks conflicting with $b$, other processes then send a \emph{1-vote} for $b$ to all  processes. 
\item Upon receiving $n-f$ 1-votes for $b$, and if they still have not seen any block conflicting with $b$, each correct process will send a \emph{2-vote} for $b$ to all others. 
\item Upon receiving $n-f$ 2-votes for $b$, a process regards $b$ as finalized. 
\end{enumerate} 
Recall that $\delta$ is the actual (unknown) bound on message delays after GST. If the new transaction block $b$ is created at time $t>$GST, then the procedure above causes all correct processes to regard $b$ as finalized by time $t+3\delta$. 

Which blocks should a new transaction block $b$ point to? For the sake of concreteness, let us specify that if there is a \emph{sole tip} amongst the blocks received by $p_i$, i.e.\ if there exists a unique block $b'$ amongst those received by $p_i$ which observes all other blocks received by $p_i$, then $p_i$ should have $b$ point to $b'$. To integrate with our approach to the `high throughput mode', we also require that $b$ should point to the last transaction block created by $p_i$. Generally, we will only require transaction blocks to point to at most two previous blocks. This avoids the downside of many DAG-based protocols that all blocks require $O(n)$ pointers to previous blocks. 

\vspace{0.2cm} 
\noindent \textbf{Moving to high throughput}. When conflicting transaction blocks are produced, we need a method for ordering them. The approach we take is to use \emph{leaders}, who produce a second type of block, called \emph{leader} blocks. These leader blocks are used to specify the required total ordering. 

\vspace{0.1cm} 
\noindent \emph{Views}. In more detail, the instructions for the protocol are divided into \emph{views}, each with a distinct leader. If a particular view is operating in `low throughput' mode and conflicting blocks are produced, then some time may pass during which a new transaction block fails to be finalized. In this case, correct processes will complain, by sending messages indicating that they wish to move to the next view. Once processes enter the next view, the leader of that view will then continue to produce leader blocks so long as the protocol remains in high throughput mode. Each of these leader blocks will point to all \emph{tips} (i.e. all blocks which are not observed by any others) seen by the leader, and will suffice to specify a total ordering on the blocks they observe. 

\vspace{0.1cm} 
\noindent \emph{The two phases of a view}. Each view is thus of potentially unbounded length and consists of two phases. During the first phase, the protocol is in high throughput mode, and is essentially the same as Autobahn.\footnote{We note that Autobahn includes the option of various optimisations (with corresponding tradeoffs) that can be used to further reduce latency in certain `good' scenarios (where all processes act correctly, for example). For the sake of simplicity we do not include these optimisations in our formal description of Morpheus in Section \ref{formalspec}, but Section \ref{metrics} discusses those options.} Processes produce transaction blocks, each of which just points to their last produced transaction block. Processes do not send 1 or 2-votes for transaction blocks during this phase, but rather vote for leader blocks, which, when finalized, suffice to specify the required total ordering on transactions. Leader blocks are finalized as in PBFT, after two rounds of voting. If a time is reached after which transaction blocks arrive infrequently enough that leader blocks are no longer required, then the view enters a second phase, during which processes vote on transaction blocks and attempt to finalize them without the use of a leader.

\vspace{0.2cm} 
\noindent \textbf{How to produce the total ordering}. For protocols in which each block points to a single precedessor, the total ordering of transactions specified by a finalized block $b$ is clear: the ordering on transactions is just that inherited by the sequence of blocks below $b$ and the transactions they contain. In a context where each block may point to multiple others, however, we have extra work to do to specify the required total ordering on transactions. The approach we take is similar to many DAG-based protocols (e.g.\ \cite{keidar2022cordial}). 
 Given any sequence of blocks $S$, we let $\mathtt{Tr}(S)$ be the corresponding sequence of transactions, i.e. if $b_1,\dots,b_k$ is the subsequence of $S$ consisting of the transaction blocks in $S$, then $\mathtt{Tr}(B)$ is $b_1.\text{Tr} \ast b_2.\text{Tr} \ast \cdots \ast b_k.\text{Tr}$, where $\ast$ denotes concatenation, and where $b.\text{Tr}$ is the sequence of transactions in $b$. We suppose given $\tau^{\dagger}$ such that, for any set of blocks $B$, $\tau^{\dagger}(B)$ is a sequence of blocks that contains each block in $B$ precisely once, and which respects the observes relation: if $b,b'\in B$ and $b'$ observes  $b$, then $b$ appears before $b'$ in $\tau^{\dagger}(B)$.  Each transaction/leader block $b$ will contain $q$ which is a 1-Quorum-Certificate (1-QC), i.e., a threshold signature formed from $n-f$ 1-votes,  for some previous block: this will be recorded as the value $b.\text{1-QC}=q$, while, if $q$ is a 1-QC for $b'$, then we set $q.\text{b}=b'$. QCs are ordered first by the view of the block to which they correspond, then by the type of the block (leader or transaction, with the latter being greater), and then by the height of the block.  We then define $\tau(b)$ by recursion: 
\begin{itemize} 
\item $\tau(b_g)=b_g$. 
\item If $b\neq b_g$, then let $q=b.\text{1-QC}$ and set $b'=q.\text{b}$. Then $\tau(b)= \tau(b') \ast \tau^{\dagger}([b]-[b'])$. 
\end{itemize} 

\begin{figure}
    \centering
    \includegraphics[width=0.7\linewidth]{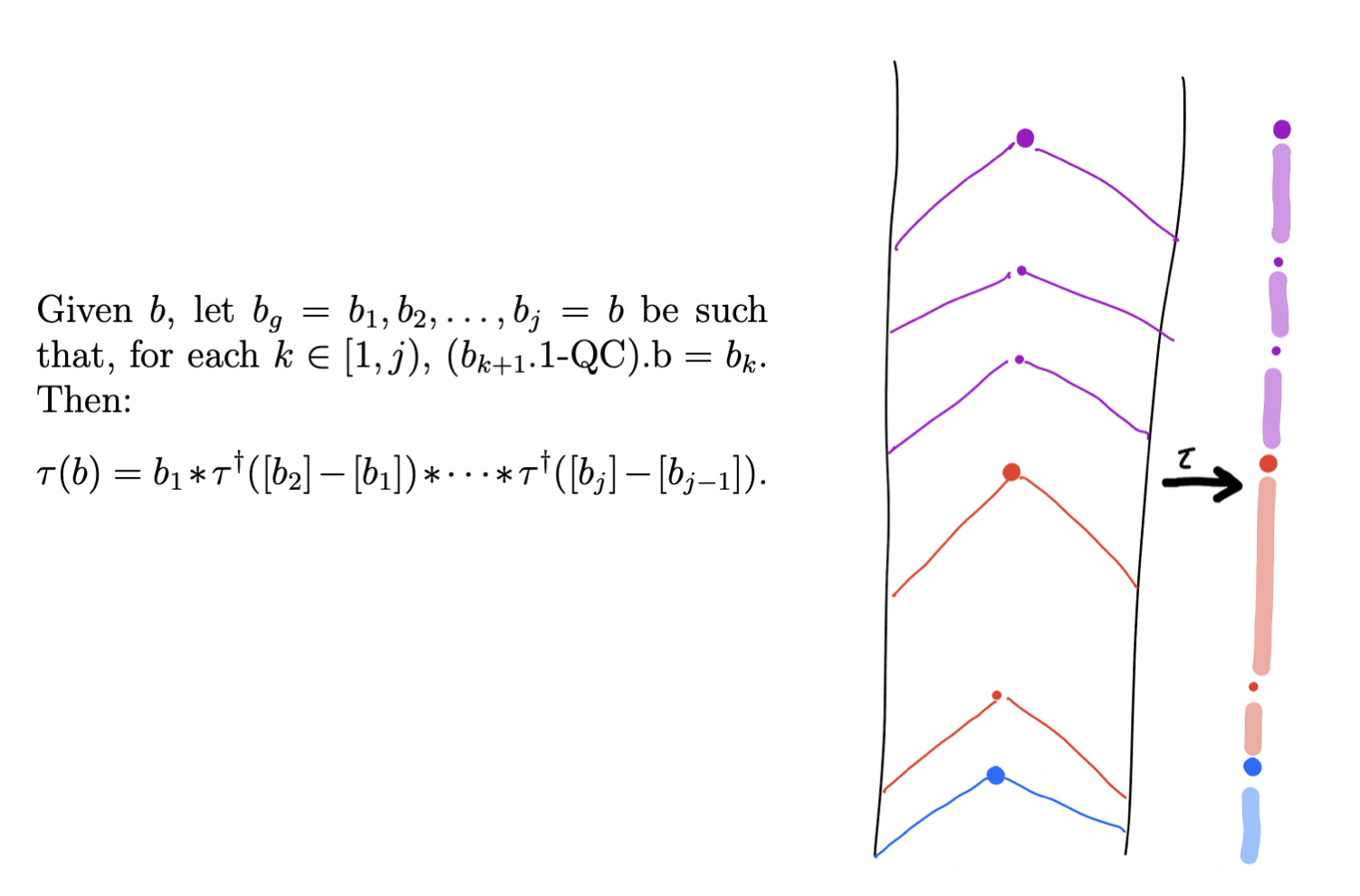}
    \caption{Specifying $\tau$ to produce the total ordering}
    \label{fig:enter-label}
\end{figure}

\vspace{0.5cm} 

\noindent   Given any set of messages $M$, let $M'$ be the largest set of blocks in $M$ that is \emph{downward closed}, i.e. such that if $b\in M'$ and $b$ observes $b'$, then $b'\in M'$. Let $q$ be a maximal  2-QC  in $M$ such that $q.\text{b}\in M'$, and set $b=q.\text{b}$, or if there is no such 2-QC  in $M$, set $b=b_g$. 
We define $\mathcal{F}(M)$ to be $\mathtt{Tr}(\tau(b))$.

\vspace{0.2cm} 
\noindent \textbf{Maintaining consistency}. Consistency is formally established in Section \ref{cl}, and uses a combination of techniques from PBFT, Tendermint, and previous DAG-based protocols.  Roughly, the argument is as follows. When the protocol moves to a new view, consistency will be maintained using the same technique as in PBFT. Upon entering the view, each process sends a `new-view' message to the leader, specifying the greatest 1-QC they have seen. Upon producing a first leader block $b$ for the view, the leader must then justify the choice of $b.\text{1-QC}$ by listing new-view messages signed by $n-f$ distinct processes in $\Pi$. The value $b.\text{1-QC}$ must be greater than or equal to all 1-QCs specified in those new-view messages. If any previous block $b'$ has received a 2-QC, then at least $f+1$ correct processes must have seen a 1-QC for $b'$, meaning that  $b.\text{1-QC}$ must be greater than or equal to that 1-QC.  Subsequent leader blocks $b''$ for the view just set $b''.\text{1-QC}$ to be a 1-QC for the previous leader block. 

To maintain consistency between finalized transaction blocks and between leader and transaction blocks within a single view, we also have each transaction block specify $q$ which is 1-QC for some previous block. Correct processes will not vote for the transaction block unless $q$ is greater than or equal to any 1-QC they have previously received. 

Overall, the result of these considerations is that, if two blocks $b$ and $b'$ receive 2-QCs $q$ and $q'$ respectively, with $q$ greater than $q'$, then the iteration specifying $\tau(b)$ (as detailed above) proceeds via $b'$, so that $\tau(b)$ extends $\tau(b')$.

\vspace{0.2cm} 
\noindent \textbf{0-votes}. While operating in low throughput, a 1-QC for a block $b$ suffices to ensure both data availability, i.e.\ that some correct process has received the block,  and \emph{non-equivocation}, i.e.\ two conflicting blocks cannot both receive 1-QCs. When operating in high throughput, however, transaction blocks will not receive 1 or 2-votes. In this context, we still wish to ensure data availability. It is also useful to ensure that each individual process does not produce transaction blocks that conflict with each other, so as to bound the number of tips that may be created. To this end, we make use of \emph{0-votes}, which may be regarded as weaker than standard votes for a block:

\begin{enumerate} 
\item Upon having a new transaction block $b$ to issue, a process $p_i$ will send $b$ to all  processes. 
\item If the block is properly formed,  and if other processes have not seen $p_i$ produce any transaction blocks conflicting with $b$, then they will send a \emph{0-vote} for $b$ back to $p_i$. Note that 0-votes are sent only to the block creator, rather than to all processes. 
\item Upon receiving $n-f$ 0-votes for $b$, $p_i$ will then form a 0-QC for $b$ and send this to all processes. 
\end{enumerate} 
\noindent When a block $b'$ wishes to point to $b$, it will include a $z$-QC for  $b$ (for some $z\in \{ 0,1,2 \}$). As a consequence, any process will be able to check that $b'$ is valid/properly formed without actually receiving the blocks that $b'$ points to:  the existence of QCs for those blocks suffices to ensure that they are properly formed (and that at least one correct process has those blocks), and other requirements for the validity of $b'$ can be checked by direct inspection. For this to work, votes (and QCs) must specify certain properties of the block beyond its hash, such as the height of the block and the block creator. The details are given in Section \ref{formalspec}.

  \section{Morpheus: the formal specification}  \label{formalspec} 
 The pseudocode uses a number of local variables, functions, objects and procedures, detailed below.  In what follows, we suppose that, when a correct process sends a message to `all processes', it regards that message as immediately received by itself. All messages are signed by the sender. For any variable $x$, we write $x\downarrow$ to denote that $x$ is defined, and $x\uparrow$ to denote that $x$ is undefined. Table 1 lists all message types. 
 
\begin{table}[h!]
  \begin{center}
  %  \caption{Your first table.}
    \label{tab:table1} 
    \begin{tabular}{l|l} % <-- Alignments: 1st column left, 2nd middle and 3rd right, with vertical lines in between
      \textbf{Message type} & \textbf{Description} \\
      \hline
      \textbf{Blocks} &  \\
      Genesis block & Unique block of height 0 \\
      Transaction blocks & Contain transactions \\
      Leader blocks & Used to totally order transaction blocks\\
            \hline
               \textbf{Votes and QCs} &  \\
                     0-votes & Guarantee data availability and non-equivocation in high throughput\\
                     1-votes & Sent during 1st round of voting on a block \\
                         2-votes & Sent during 2nd round of voting on a block \\
    
                      $z$-QC, $z\in \{ 0,1,2 \}$ & formed from $n-f$ $z$-votes \\
                             \hline
                           \textbf{View messages} &  \\
                            End-view messages & Indicate wish to enter next view\\
                            $(v+1)$-certificate & Formed from $f+1$ end-view $v$ messages\\
                            View $v$ message & Sent to the leader at start of view $v$\\

    \end{tabular}
        \caption{Message types.}

  \end{center}
\end{table}

 \vspace{0.1cm} 
 \noindent \textbf{The genesis block}.  There exists a unique \emph{genesis block}, denoted $b_g$. For any block $b$, $b.\text{type}$ specifies the type of the block $b$, $b.\text{view}$ is the \emph{view} corresponding to the block, $b.\text{h}$ specifies the \emph{height} of the block, $b.\text{auth}$ is the block creator, and $b.\text{slot}$  specifies the \emph{slot} corresponding to the block. For $b_g$, we set: 
 \begin{itemize} 
 \item  $b_g.\text{type}=\text{gen}$, $b_g.\text{view}=-1$, $b_g.\text{h}=0$, $b_g.\text{auth}=\bot$, $b_g.\text{slot}=0$. 
% \item $b_g.\text{view}=0$. 
% \item  $b_g.\text{h}=0$.
 \end{itemize}  
 
% \noindent \hrulefill
% 
% \vspace{0.1cm} 
% \noindent \emph{A comment on the use of slots}. Each block will either be the genesis block, a \emph{transaction} block, or a \emph{leader} block. If $p_i\in \Pi$ is correct then, for $s\in \mathbb{N}_{\geq 0}$,  $p_i$ will produce a single  transaction block $b$ with $b.\text{slot}=s$ before producing any transaction  block $b'$ with $b'.\text{slot}=s+1$. Similarly, if $p_i\in \Pi$ is correct then, for $s\in \mathbb{N}_{\geq 0}$,  $p_i$ will produce a single leader  block $b$ with $b.\text{slot}=s$ before producing any leader block $b'$ with $b'.\text{slot}=s+1$.
%
%
%
% \noindent \hrulefill

 \begin{tcolorbox}[colback=gray!5!white,colframe=black!75!black,top=4pt,bottom=4pt]
 
 \noindent \emph{A comment on the use of slots}. Each block will either be the genesis block, a \emph{transaction} block, or a \emph{leader} block. If $p_i\in \Pi$ is correct then, for $s\in \mathbb{N}_{\geq 0}$,  $p_i$ will produce a single  transaction block $b$ with $b.\text{slot}=s$ before producing any transaction  block $b'$ with $b'.\text{slot}=s+1$. Similarly, if $p_i\in \Pi$ is correct then, for $s\in \mathbb{N}_{\geq 0}$,  $p_i$ will produce a single leader  block $b$ with $b.\text{slot}=s$ before producing any leader block $b'$ with $b'.\text{slot}=s+1$.

 \end{tcolorbox} 
 
%  \vspace{0.1cm} 
% \noindent \textbf{Approval-votes}. An \emph{approval-vote} for the block $b$ is a message of the form $(\text{app}, b.\text{type},b.\text{view},$ $b.\text{h},b.\text{auth}, b.\text{slot}, H(b))$ signed by some process in $\Pi$. The reason approval-votes include more information than just the hash of the block is explained in Section \ref{intuition}. 
%By an \emph{approval-quorum} for $b$, we mean $n-f$ approval-votes for $b$, each signed by a different process in $\Pi$. By an \emph{approval-certificate} for $b$, we mean the  message $m= (\text{app}, b.\text{type},b.\text{view},$ $b.\text{h}, b.\text{auth}, b.\text{slot}, H(b))$ together with a threshold signature for $m$, formed using the threshold signature scheme from an approval-quorum for $b$. 

   \vspace{0.1cm} 
 \noindent \textbf{$z$-votes}. For $z\in \{ 0, 1, 2 \}$, a \emph{$z$-vote} for the block $b$ is a message of the form $(z, b.\text{type},b.\text{view},$ $b.\text{h},b.\text{auth}, b.\text{slot}, H(b))$, signed by some process in $\Pi$. The reason votes include more information than just the hash of the block is explained in Section \ref{intuition}.  A \emph{$z$-quorum} for $b$ is a set of $n-f$ $z$-votes for $b$, each signed by a different process in $\Pi$. A \emph{$z$-QC} for $b$ is the message  $m=(z, b.\text{type},b.\text{view},$ $b.\text{h}, b.\text{auth}, b.\text{slot},H(b))$ together with a threshold signature for $m$, formed  from a $z$-quorum for $b$ using the threshold signature scheme. 
 
    \vspace{0.1cm} 
 \noindent \textbf{QCs}. By a \emph{QC} for the block $b$, we mean a $z$-QC for $b$, for some $z\in \{ 0,1,2 \}$. 
% \begin{itemize} 
% \item A set of $n-f$ approval-votes for $b$, each signed by a different process in $\Pi$, or; 
% \item A $z$-QC for $b$, where $z\in \{ 1, 2 \}$. 
% \end{itemize} 
% By an \emph{approval-certificate} for $b$, we mean either: 
%  \begin{itemize} 
% \item The message $m= (\text{app}, b.\text{type},b.\text{view},$ $b.\text{h}, b.\text{auth}, b.\text{slot}, H(b))$ together with a threshold signature for $m$, where the latter is formed from $n-f$ signatures for $m$, each by a different process in $\Pi$, or; 
% \item A $z$-QC-certificate for $b$, where $z\in \{ 1, 2 \}$. 
% \end{itemize} 
 \noindent If $q$ is a $z$-QC for $b$, then we set $q.\text{b}=b$, $q.\text{z}=z$, $q.\text{type}=b.\text{type}$, $q.\text{view}=b.\text{view},$ $q.\text{h}=b.\text{h}$, $q.\text{auth}=b.\text{auth}$, $q.\text{slot}=b.\text{slot}$. We  define a preordering $\leq$ on QCs as follows: QCs are preordered first by view, then by type with $\text{lead}<\text{Tr}$,  and then by height.\footnote{For the sake of completeness, if $q.\text{view}=q'.\text{view}$,  $q.\text{type}=q'.\text{type}$, and  $q.\text{h}=q'.\text{h}$, then we set $q\leq q'$ and $q'\leq q$. We will show that, in this case, $q.\text{b}=q'.\text{b}$.}

 \vspace{0.1cm} 
 \noindent \textbf{The variable} $M_i$. Each process $p_i$ maintains a local variable $M_i$, which is automatically updated and specifies the set of all received messages. Initially, $M_i$ contains $b_g$ and a 1-QC  for $b_g$.

  \vspace{0.1cm} 
 \noindent \textbf{Transaction blocks}. Each \emph{transaction block} $b$ is entirely specified by the following values: 
  \begin{itemize} 
   \item  $b.\text{type}=\text{Tr}$,  $b.\text{view}=v\in \mathbb{N}_{\geq 0}$,  $b.\text{h}=h\in \mathbb{N}_{> 0}$, $b.\text{slot}=s\in \mathbb{N}_{\geq 0}$. 
 %  \item $b.\text{vslot}=\bot$.
% \item $b.\text{view}=v\in \mathbb{N}_{\geq 0}$. 
% \item  $b.\text{h}=h\in \mathbb{N}_{> 0}$.
 \item $b.\text{auth}\in \Pi$: the block creator. 
 \item $b.\text{Tr}$: a sequence of transactions. 
 \item $b.\text{prev}$: a non-empty set of QCs for blocks of height $<h$.  
 \item $b.\text{1-QC}$: a 1-QC for a block of height $<h$.
 \end{itemize}  
 \noindent If $b.\text{prev}$ contains a QC for $b'$, then we say that $b$ \emph{points to} $b'$. For $b$ to be \emph{valid}, we require that it is of the form above and: 
 \begin{enumerate} 
 \item $b$ is signed by $b.\text{auth}$. 
% \item $b$ points to at most two blocks. 
 \item If $s>0$, $b$ points to $b'$ with $b'.\text{type}=\text{Tr}$,  $b'.\text{auth}=b.\text{auth}$ and $b'.\text{slot}=s-1$. 
% \item $b.\text{prev}$ contains at most two approval-certificates, one of which is for $b'$ with $b'.\text{auth}=b.\text{auth}$ and $b'.\text{slot}=s-1$. 
 \item If $b$ points to $b'$, then $b'.\text{view}\leq b.\text{view}$. 
 \item If $h'= \text{max} \{ b'.\text{h}:\ b \text{ points to }b'\}$, then $h=h'+1$. 
 \end{enumerate} 
 \noindent We suppose correct processes ignore transaction blocks that are not valid. In what follows we therefore adopt the convention that, by a `transaction block', we mean a `valid transaction block'.

    \begin{tcolorbox}[colback=gray!5!white,colframe=black!75!black,top=4pt,bottom=4pt]
 
 \noindent \emph{A comment on transaction blocks}. During periods of high throughput, a transaction block produced by $p_i$ for slot $s$ will just point to $p_i$'s transaction block for slot $s-1$. During periods of low throughput, if there is a unique block $b'$ received by $p_i$ that does not conflict with any other block received by $p_i$, any transaction block $b$ produced by  $p_i$ will also point to $b'$ (so that $b$ does not conflict with $b'$). 
 
 The use of $b.\text{1-QC}$ is as follows: once correct $p_i$ sees a 1-QC $q$, it will not vote for any transaction block $b$ unless $b.\text{1-QC}$ is greater than or equal to $q$. Ultimately, this will be used to argue that consistency is satisfied. 

 \end{tcolorbox} 
 
%   \noindent \hrulefill
% 
% \vspace{0.1cm} 
% \noindent \emph{A comment on transaction blocks}. During periods of high throughput, a transaction block produced by $p_i$ for slot $s$ will just point to $p_i$'s transaction block for slot $s-1$. During periods of low throughput, if there is a unique block $b'$ received by $p_i$ that does not conflict with any other block received by $p_i$ (with an availability certificate), any transaction block $b$ produced by (correct) $p_i$ will also point to $b'$ (so that $b$ does not conflict with $b'$). 
% 
% The use of $b.\text{1-QC}$ is as follows: once correct $p_i$ sees $q$ which is a 1-QC-certificate for some block $b'$, it will not vote for any transaction block $b$ unless $b.\text{1-QC}$ is greater than or equal to $q$. Ultimately, this will be used to argue that consistency is satisfied. 
% 
%    \noindent \hrulefill
    
    \vspace{0.1cm} 
    \noindent \textbf{When blocks observe each other}. The genesis block \emph{observes} only itself. Any other block $b$ observes itself and all those blocks observed by blocks that $b$ points to. If two blocks do not observe each other, then they \emph{conflict}. We write $[b]$ to denote the set of all blocks observed by $b$. 
    
        \vspace{0.1cm} 
    \noindent \textbf{The leader of view $v$}. The leader of view $v$, denoted $\mathtt{lead}(v)$, is process $p_i$, where $i=v \text{ mod } n$. 
 
   \vspace{0.1cm} 
 \noindent \textbf{End-view messages}.  If process $p_i$ sees insufficient progress during view $v$, it may send an end-view $v$ message of the form $(v)$, signed by $p_i$. 
By a \emph{quorum} of end-view $v$ messages, we mean a set of $f+1$ end-view $v$ messages, each signed by a different process in $\Pi$. If $p_i$ receives a quorum of end-view $v$ messages  before entering view $v+1$, it will combine them (using the threshold signature scheme) to form a $(v+1)$-certificate. Upon first seeing a $(v+1)$-certificate, $p_i$ will send this certificate to all processes and enter view $v+1$. This ensures that, if  some correct process is the first to enter view $v+1$ after GST,  all correct processes enter that view (or a later view) within time $\Delta$. 
 
    \vspace{0.1cm} 
 \noindent \textbf{View $v$ messages}. When $p_i$ enters view $v$, it will send to $\mathtt{lead}(v)$ a  view $v$ message of the form $(v,q)$, signed by $p_i$, where $q$ is a maximal amongst 1-QCs seen by $p_i$. We say that $q$ is the 1-QC \emph{corresponding} to the  view $v$ message  $(v,q)$.

     \begin{tcolorbox}[colback=gray!5!white,colframe=black!75!black,top=4pt,bottom=4pt]
 
 \noindent \emph{A comment on view $v$ messages}. 
  The use of view $v$ messages is to carry out view changes in the same manner as PBFT. When producing the first leader block $b$ of the view, the leader must include a set of $n-f$ view $v$ messages, which act as a \emph{justification} for the block proposal: the value $b.\text{1-QC}$ must be greater than or equal all 1-QCs corresponding to  those $n-f$ view $v$ messages. For each subsequent leader block $b'$ produced in the view, $b'.\text{1-QC}$  must be a 1-QC  for the previous leader block (i.e.,\ that for the previous slot). The argument for consistency will thus employ some of the same methods as are used to argue consistency for PBFT.

 \end{tcolorbox} 
 
%    \noindent \hrulefill
% 
% \vspace{0.1cm} 
% \noindent \emph{A comment on view $v$ messages}. 
%  The use of view $v$ messages is to carry out view changes in the same manner as PBFT \cite{}. When producing the first leader block $b$ of the view, the leader must include a set of $n-f$ view $v$ messages, which act as a \emph{justification} for the block proposal: the value $b.\text{1-QC}$ must be greater than or equal all 1-QC-certificates corresponding to  those $n-f$ view $v$ messages. For each subsequent leader block $b'$ produced in the view, $b'.\text{1-QC}$  must be a 1-QC-certificate for the previous leader block (i.e.,\ that for the previous slot). The argument for consistency will thus employ some of the same methods as are used to argue consistency for PBFT. 
%
%    \noindent \hrulefill

   \vspace{0.1cm} 
 \noindent \textbf{Leader blocks}. Each \emph{leader block} $b$ is entirely specified by the following values: 
  \begin{itemize} 
   \item  $b.\text{type}=\text{lead}$,  $b.\text{view}=v\in \mathbb{N}_{\geq 0}$,  $b.\text{h}=h\in \mathbb{N}_{> 0}$, $b.\text{slot}=s\in \mathbb{N}_{\geq 0}$. 
 %  \item $b.\text{vslot}=x\in \mathbb{N}_{\geq 0}$.
% \item $b.\text{view}=v\in \mathbb{N}_{\geq 0}$. 
% \item  $b.\text{h}=h\in \mathbb{N}_{> 0}$.
 \item $b.\text{auth}\in \Pi$: the block creator. 
 \item $b.\text{prev}$: a non-empty set of QCs for blocks of height $<h$.  
 \item $b.\text{1-QC}$: a 1-QC  for a block of height $<h$.
 \item $b.\text{just}$: a (possibly empty) set of view $v$ messages. 
 \end{itemize}  
 \noindent As for transaction blocks, if $b.\text{prev}$ contains a QC for $b'$, then we say that $b$ \emph{points to} $b'$. For $b$ to be \emph{valid}, we require that it is of the form described above and: 
 \begin{enumerate} 
 \item $b$ is signed by $b.\text{auth}$ and $b.\text{auth}=\mathtt{lead}(v)$. 
  \item If $b$ points to $b'$, then $b'.\text{view}\leq b.\text{view}$. 
   \item If $h'= \text{max} \{ b'.\text{h}:\ b \text{ points to }b'\}$, then $h=h'+1$. 
% \item $b$ points to at most two blocks. 
 \item If $s>0$, $b$ points to a unique $b^*$ with $b^*.\text{type}=\text{lead}$,  $b^*.\text{auth}=b.\text{auth}$ and $b^*.\text{slot}=s-1$. 
% \item $b.\text{prev}$ contains at most two approval-certificates, one of which is for $b'$ with $b'.\text{auth}=b.\text{auth}$ and $b'.\text{slot}=s-1$. 
 \item If $s=0$ or $b^*.\text{view}<v$, then $b.\text{just}$ contains $n-f$ view $v$ messages, each signed by a different process in $\Pi$. This set of messages is called a \emph{justification} for the block. 
 \item If $s=0$ or $b^*.\text{view}<v$, then $b.\text{1-QC}$ is greater than or equal to all 1-QCs corresponding to view $v$ messages in $b.\text{just}$. 
 \item If $s>0$ and $b^*.\text{view}=v$, then $b.\text{1-QC}$ is a 1-QC for $b^*$. 
 \end{enumerate} 
 \noindent As with transaction blocks, we suppose correct processes ignore leader blocks that are not valid. In what follows we therefore adopt the convention that, by a `leader block', we mean a `valid leader block'.

      \begin{tcolorbox}[colback=gray!5!white,colframe=black!75!black,top=4pt,bottom=4pt]
 
 \noindent \emph{A comment on leader blocks}. The conditions for validity above are just those required to carry out a PBFT-style approach to view changes (as discussed previously). 
 The first leader block of the view must include a justification for the block proposal (to guarantee consistency). Subsequent leader blocks in the view simply include a 1-QC for the previous leader block (i.e.,\ that for the previous slot).

 \end{tcolorbox} 
 
%     \noindent \hrulefill
% 
% \vspace{0.1cm} 
% \noindent \emph{A comment on leader blocks}. The conditions for validity above are just those required to carry out a PBFT-style approach to view changes (as discussed previously). 
% The first leader block of the view must include a justification for the block proposal (to guarantee consistency). Subsequent leader blocks in the view simply include a 1-QC-certificate for the previous leader block (i.e.,\ that for the previous slot). 
%
%
%    \noindent \hrulefill

   \vspace{0.1cm} 
 \noindent \textbf{The variable} $Q_i$.   Each process $p_i$ maintains a local variable $Q_i$, which is automatically updated and, for each $z\in \{ 0,1,2 \}$, stores at most one $z$-QC for  each block: For $z\in \{ 0,1,2 \}$, if $p_i$ receives\footnote{Here, we include the possibility that $p_i$ receives the $z$-QC inside a message, such as in $b'.\text{prev}$ for a received block $b'$} a $z$-quorum or a $z$-QC for  $b$, and if $Q_i$ does not contain a $z$-QC for $b$, then $p_i$ automatically enumerates a $z$-QC for $b$ into $Q_i$ (either the $z$-QC received, or one formed from the $z$-quorum received).
%  \begin{itemize} 
% \item If $p_i$ receives  an  $q$  for a block $b$ (possibly inside a message, such as included in $b'.\text{prev}$ for a received block $b'$), and if $Q_i$ does not contain an approval-certificate for $b$, then $p_i$ automatically enumerates $q$  into $Q_i$. 
%   \item For $z\in \{ 1,2 \}$, if $p_i$ receives a $z$-quorum or a $z$-QC for  $b$, and if $Q_i$ does not contain a $z$-QC for $b$, then $p_i$ automatically enumerates a $z$-QC for $b$ into $Q_i$ (either the $z$-QC received, or one formed from the $z$-quorum received).
% \end{itemize} 
% 
 
We define the `observes' relation $\succeq$ on $Q_i$ to be the minimal preordering satisfying (transitivity and): 
 \begin{itemize}
 \item If $q,q'\in Q_i$, $q.\text{type}=q'.\text{type}$, $q.\text{auth}=q'.\text{auth}$ and $q.\text{slot}>  q'.\text{slot}$, then $q \succeq q'$. 
 \item  If $q,q'\in Q_i$, $q.\text{type}=q'.\text{type}$, $q.\text{auth}=q'.\text{auth}$, $q.\text{slot}=q'.\text{slot}$, and $q.\text{z}\geq q'.\text{z}$, then $q \succeq q'$. 
 \item If $q,q'\in Q_i$, $q.\text{b}=b$, $q'.\text{b}=b'$, $b\in M_i$ and $b$ points to $b'$, then $q \succeq q'$. 
 \end{itemize}  
 We note that the observes relation $\succeq$ depends on $Q_i$ and $M_i$, and is stronger than the preordering $\geq$ we defined on $z$-QCs previously, in the following sense: if $q$ and $q'$ are $z$-QCs with $q\succeq q'$, then $q\geq q'$, while the converse may not hold. When we refer to the `greatest' QC in a given set, or a `maximal' QC in a given set, this is with reference to the $\geq$ preordering, unless explicitly stated otherwise. If $q.\text{type}=q'.\text{type}$, $q.\text{auth}=q'.\text{auth}$ and $q.\text{slot}= q'.\text{slot}$, then it will follow that $q.\text{b}=q'.\text{b}$.

      \begin{tcolorbox}[colback=gray!5!white,colframe=black!75!black,top=4pt,bottom=4pt]
 
 \noindent \emph{A comment on the observes relation on $Q_i$}. When $p_i$ receives $q,q'\in Q_i$, it may not be immediately apparent whether $q.\text{b}$ observes $q'.\text{b}$. The observes relation defined on $Q_i$ above is essentially that part of the observes relation on blocks that $p_i$ can testify to, given the messages it has received (while also distinguishing the `level' of the QC).

 \end{tcolorbox} 
 
%      \noindent \hrulefill
% 
% \vspace{0.1cm} 
% \noindent \emph{A comment on the observes relation on $Q_i$}. When $p_i$ receives $q,q'\in Q_i$, it may not be immediately apparent whether $q.\text{b}$ observes $q'.\text{b}$. The observes relation defined on $Q_i$ above is essentially that part of the observes relation on blocks that $p_i$ can testify to, given the messages it has received. 
% 
%       \noindent \hrulefill
 
    \vspace{0.1cm} 
 \noindent \textbf{The tips of} $Q_i$. The \emph{tips} of $Q_i$ are those $q\in Q_i$ such that there does not exist $q'\in Q_i$ with $q' \succ q$ (i.e. $q'\succeq q$ and $q \not \succeq q'$). The protocol ensures that $Q_i$ never contains more than $2n$ tips: The factor 2 here comes from the fact that leader blocks produced by correct $p_i$ need not observe all transaction blocks produced by $p_i$ (and vice versa).

     \vspace{0.1cm} 
 \noindent \textbf{Single tips}. We say $q\in Q_i$ is a \emph{single tip of $Q_i$} if $q\succeq q'$ for all $q'\in Q_i$. We say $b\in M_i$ is a \emph{single tip of $M_i$} if there exists $q$ which is a single tip of $Q_i$ and $b$ is the unique block in $M_i$ pointing to $q.\text{b}$. 
 
       \begin{tcolorbox}[colback=gray!5!white,colframe=black!75!black,top=4pt,bottom=4pt]
 
 \noindent \emph{A comment on single tips}. When a transaction block is a single tip of $M_i$, this will enable $p_i$ to send a 1-vote for the block. Leader blocks do not have to be single tips for correct processes to vote for them.

 \end{tcolorbox} 
 
%       \noindent \hrulefill
% 
% \vspace{0.1cm} 
% \noindent \emph{A comment on single tips}. When a transaction block is a single tip of $M_i$, this will enable $p_i$ to vote for the block. Leader blocks do not have to be single tips for correct processes to vote for them. 
%     
%       \noindent \hrulefill
 
%       \vspace{0.1cm} 
% \noindent \textbf{The $\mathtt{approved}$ function}. For each $i,j,s$ and $x\in \{ \text{lead}, \text{Tr} \}$, the value $\mathtt{approved}_i(x,s,p_j)$ is initially 0. When $p_i$ sends an approval-vote for a block $b$ with $b.\text{type}=x$, $b.\text{auth}=p_j$, and $b.\text{slot}=s$, it sets $\mathtt{approved}_i(x,s,p_j):=1$. Once this value is set to 1, $p_i$ will not send an approval-vote for any block $b'$ with   $b'.\text{type}=x$, $b'.\text{auth}=p_j$, and $b'.\text{slot}=s$. 

      \vspace{0.1cm} 
 \noindent \textbf{The $\mathtt{voted}$ function}. For each $i,j,s$, $z\in \{ 0,1,2 \}$ and $x\in \{ \text{lead}, \text{Tr} \}$, the value $\mathtt{voted}_i(z,x,s,p_j)$ is initially 0. When $p_i$ sends a $z$-vote for a block $b$ with $b.\text{type}=x$, $b.\text{auth}=p_j$, and $b.\text{slot}=s$, it sets $\mathtt{voted}_i(z,x,s,p_j):=1$. Once this value is set to 1, $p_i$ will not send a $z$-vote for any block $b'$ with   $b'.\text{type}=x$, $b'.\text{auth}=p_j$, and $b'.\text{slot}=s$.

       \vspace{0.1cm} 
 \noindent \textbf{The phase during the view}. For each $i$ and $v$, the value $\mathtt{phase}_i(v)$ is initially 0. Once $p_i$ votes for a transaction block during view $v$, it will set 
 $\mathtt{phase}_i(v):=1$, and will then not vote for leader blocks within view $v$.

        \begin{tcolorbox}[colback=gray!5!white,colframe=black!75!black,top=4pt,bottom=4pt]
 
 \noindent \emph{A comment on the phase during a view}. As noted previously, each view can be thought of as consisting of two phases. Initially, the leader is responsible for finalizing transactions. If, after some time, the protocol enters a period of low throughput, then the leader will stop producing leader blocks, and transactions blocks can then be finalized directly. Once a process votes for a transaction block, it may be considered as having entered the low throughput phase of the view. The requirement that it should not then vote for subsequent leader blocks in the view is made so as to ensure consistency between finalized leader blocks and transaction blocks within the view.

 \end{tcolorbox} 
 
%         \noindent \hrulefill
% 
% \vspace{0.1cm} 
% \noindent \emph{A comment on the phase during a view}. As noted previously, each view can be thought of as consisting of two phases. Initially, the leader is responsible for finalizing transactions. If, after some time, the protocol enters a period of low throughput, then the leader will stop producing leader blocks, and transactions blocks can then be finalized directly. Once a process votes for a transaction block, it may be considered as having entered the low throughput phase of the view. The requirement that it should not then vote for subsequent leader blocks in the view is made so as to ensure consistency between finalized leader blocks and transaction blocks within the view.  
%      
%       \noindent \hrulefill

        \vspace{0.1cm} 
 \noindent \textbf{When blocks are final}. Process $p_i$ regards $q\in Q_i$ (and $q.\text{b}$) as \emph{final} if there exists $q'\in Q_i$ such that $q'\succeq q$ and $q$ is a 2-QC  (for any block). 
 
           \vspace{0.1cm} 
 \noindent \textbf{The function $\mathcal{F}$}. This is defined exactly as specified in Section \ref{intuition}.

         \vspace{0.1cm} 
 \noindent \textbf{The variables $\mathtt{view}_i$ and $\mathtt{slot}_i(x)$ for }$x\in \{ \text{lead}, \text{Tr} \}$. These record the present view and slot numbers for $p_i$. 

         \vspace{0.1cm} 
 \noindent \textbf{The $\mathtt{PayloadReady}_i$ function}. We remain agnostic as to how frequently processes should produce transaction blocks, i.e.\ as to whether processes should produce transaction blocks immediately upon having  new transactions to process, or wait until they have a set of new transactions of at least a certain size. We suppose simply that: 
 \begin{itemize} 
 \item  Extraneous to the explicit instructions of the protocol, $\mathtt{PayloadReady}_i$ may be set to 1 at some timeslots of the execution. 
 \item If $\mathtt{PayloadReady}_i=1$ and $\mathtt{slot}_i(\text{Tr})=s>0$, then there exists $q\in Q_i$ with $q.\text{auth}=p_i$, $q.\text{type}=\text{Tr}$ and $q.\text{slot}=s-1$. 
\end{itemize}  

        \begin{tcolorbox}[colback=gray!5!white,colframe=black!75!black,top=4pt,bottom=4pt]
 
 \noindent \emph{A comment on the $\mathtt{PayloadReady}_i$ function}. The second requirement above is required so that $p_i$ can ensure that the new transaction block it forms can point to its transaction block for the previous slot.

 \end{tcolorbox} 

%         \noindent \hrulefill
%  
%  \vspace{0.1cm} 
% \noindent \emph{A comment on the $\mathtt{PayloadReady}_i$ function}. The second requirement above is required so that $p_i$ can ensure that the new transaction block it forms can point to its transaction block for the previous slot.
% 
%          \noindent \hrulefill

          \vspace{0.1cm} 
 \noindent \textbf{The procedure $\mathtt{MakeTrBlock}_i$}. When $p_i$ wishes to form a new transaction block $b$, it will run this procedure, by executing the following instructions: 
 \begin{enumerate} 
 \item Set   $b.\text{type}:=\text{Tr}$,  $b.\text{auth}:=p_i$, $b.\text{view}:=\mathtt{view}_i$,  $b.\text{slot}:=\mathtt{slot}_i(\text{Tr})$. 
 \item Let $s:=\mathtt{slot}_i(\text{Tr})$. If $s>0$,  then let $q_1\in Q_i$ be such that  $q_1.\text{auth}=p_i$, $q_1.\text{type}=\text{Tr}$ and $q_1.\text{slot}=s-1$. If $s=0$, let $q_1$ be a 1-QC for $b_g$. Initially, set $b.\text{prev}:=\{ q_1 \}$. 
 \item If there exists $q_2\in Q_i$ which is a single tip of $Q_i$, then enumerate $q_2$ into  $b.\text{prev}$. 
 \item  If $h'= \text{max} \{ q.\text{h}:\ q \in b.\text{prev} \}$, then set $b.\text{h}:=h'+1$. 
 \item Let $q$ be the  greatest 1-QC  in $Q_i$. Set $b.\text{1-QC}:=q$. 
 \item Sign $b$ with the values specified above, and send this block to all processes. 
 \item Set $\mathtt{slot}_i(\text{Tr}):=\mathtt{slot}_i(\text{Tr})+1$; 
 \end{enumerate} 
 
           \vspace{0.1cm} 
 \noindent \textbf{The boolean $\mathtt{LeaderReady}_i$}. At any time, this boolean is equal to 1 iff either of the following conditions are satisfied, setting $v=\mathtt{view}_i$: 
 \begin{enumerate} 
 \item Process $p_i$ has not yet produced a block $b$ with $b.\text{view}=v$ and $b.\text{type}=\text{lead}$, and both: 
 \begin{enumerate} 
 \item Process $p_i$ has received view $v$ messages signed by at least $n-f$ processes in $\Pi$. 
 \item $\mathtt{slot}_i(\text{lead})=0$ or $Q_i$ contains $q$ with $q.\text{auth}=p_i$, $q.\text{type}=\text{lead}$, $q.\text{slot}=\mathtt{slot}_i(\text{lead})-1$. 
 \end{enumerate}
  \item Process $p_i$ has previously produced a block $b$ with $b.\text{view}=v$ and $b.\text{type}=\text{lead}$, and $Q_i$ contains a 1-QC for  $b'$ with $b'.\text{auth}=p_i$, $b'.\text{type}=\text{lead}$, $b'.\text{slot}=\mathtt{slot}_i(\text{lead})-1$.
 \end{enumerate} 
 
         \begin{tcolorbox}[colback=gray!5!white,colframe=black!75!black,top=4pt,bottom=4pt]
 
 \noindent \emph{A comment on the boolean $\mathtt{LeaderReady}_i$}. If $p_i$ is the leader for view $v$, then before producing the first leader block of the view, it must receive view $v$ messages from $n-f$ different processes, and must also receive a QC for the last leader block it produced (if any). Before producing any subsequent leader block in the view, it must receive a 1-QC for the previous leader block.

 \end{tcolorbox} 
 
%          \noindent \hrulefill
% 
%  \vspace{0.1cm} 
% \noindent \emph{A comment on the boolean $\mathtt{LeaderReady}_i$}. If $p_i$ is the leader for view $v$, then before producing the first leader block of the view, it must receive view $v$ messages from $n-f$ different processes, and must also receive an approval certificate for the last leader block it produced (if any). Before producing any subsequent leader block in the view, it must receive a 1-QC-certificate for the previous leader block.
%  
%          \noindent \hrulefill
          
           \vspace{0.1cm} 
 \noindent \textbf{The procedure $\mathtt{MakeLeaderBlock}_i$}. When $p_i$ wishes to form a new leader block $b$, it will run this procedure, by executing the following instructions: 
 \begin{enumerate} 
 \item Set   $b.\text{type}:=\text{lead}$,  $b.\text{auth}:=p_i$, $b.\text{view}:=\mathtt{view}_i$,  $b.\text{slot}:=\mathtt{slot}_i(\text{lead})$. 
 \item Initially, set $b.\text{prev}$ to be the tips of $Q_i$. 
 \item Set $s:=\mathtt{slot}_i(\text{Tr})$ and $v:=\mathtt{view}_i$. If $s>0$,  then let $q\in Q_i$ be such that  $q.\text{auth}=p_i$, $q.\text{type}=\text{lead}$ and $q.\text{slot}=s-1$. If $b.\text{prev}$ does not already contain $q$, add $q$ to this set. 
 \item  If $h'= \text{max} \{ q.\text{h}:\ q \in b.\text{prev} \}$, then set $b.\text{h}:=h'+1$. 
 \item  If  $p_i$ has not yet produced a block $b$ with $b.\text{view}=\mathtt{view}_i$ and $b.\text{type}=\text{lead}$ then: 
 \begin{enumerate} 
 \item Set $b.\text{just}$ to be a set of view $v$ messages signed by $n-f$ processes in $\Pi$. 
 \item Set $b.\text{1-QC}$ to be a 1-QC in $Q_i$ greater than or equal to all 1-QCs corresponding to messages in $b.\text{just}$. 
 \end{enumerate} 
 \item If $p_i$ has previously  produced a block $b$ with $b.\text{view}=\mathtt{view}_i$ and $b.\text{type}=\text{lead}$ then let $q'\in Q_i$ be a 1-QC with  $q'.\text{auth}=p_i$, $q'.\text{type}=\text{lead}$ and $q'.\text{slot}=s-1$. Set $b.\text{1-QC}:=q'$ and set $b.\text{just}$ to be the empty set. 
 \item Sign $b$ with the values specified above, and send this block to all processes. 
  \item Set $\mathtt{slot}_i(\text{lead}):=\mathtt{slot}_i(\text{lead})+1$; 
 \end{enumerate}

 \vspace{0.2cm} 
 \noindent \textbf{The pseudocode}. 
 The pseudocode appears in Algorithm 1 (with local variables described first, and the main code appearing later). Section \ref{wt} gives a `pseudocode walk-through'.

\begin{algorithm} \label{pc:Snowflake}
\caption{Morpheus:  local variables for  $p_i$}
\begin{algorithmic}[1]

    \State \textbf{Local variables} 
    
   \State $M_i$, initially contains $b_g$ and a 1-QC-certificate for $b_g$   \Comment Automatically updated
   
     \State $Q_i$, initially contains 1-QC-certificate for $b_g$    \Comment Automatically updated
     
     \State $\mathtt{view}_i$, initially 0 \Comment The present view
     
          \State $\mathtt{slot}_i(x)$ for $x\in \{ \text{lead},\text{Tr} \}$, initially 0 \Comment Present slot
          
      %    \State $\mathtt{approved}_i(x,s,p_j)$ for  $x\in \{ \text{lead},\text{Tr} \}$, $s\in \mathbb{N}_{\geq 0}$, $p_j\in \Pi$, initially 0
          
        \State   $\mathtt{voted}_i(z,x,s,p_j)$ for $z\in \{0,1,2 \}$,  $x\in \{ \text{lead},\text{Tr} \}$, $s\in \mathbb{N}_{\geq 0}$, $p_j\in \Pi$, initially 0
        
        \State $\mathtt{phase}_i(v)$ for $v\in \mathbb{N}_{\geq 0}$, initially 0 \Comment The phase within the view
        
        \State  \textbf{Other procedures and functions} 
        
        \State $\mathtt{lead}(v)$ \Comment Leader of view $v$ 
        
        \State  $\mathtt{PayloadReady}_i$  \Comment Set to 1 when ready to produce transaction block
        
        \State  $\mathtt{MakeTrBlock}_i$ \Comment Sends a new transaction block to all 
        
        \State $\mathtt{LeaderReady}_i$ \Comment Indicates whether ready to produce leader block
        
        \State $\mathtt{MakeLeaderBlock}_i$ \Comment Sends a new leader block to all

\algstore{endvar}

\end{algorithmic}
\end{algorithm}

\setcounter{algorithm}{0}

\begin{algorithm} 
\caption{Morpheus: The instructions for  $p_i$ }
\begin{algorithmic}[1]

\algrestore{endvar}

\State Process $p_i$ executes the following transitions at timeslot $t$ (according to its local clock), until no further transitions apply. If multiple transitions apply simultaneously, then $p_i$ executes the first that applies, before checking whether further transitions apply, and so on.  

%\State 

 \hrule 
 
 \State \Comment Update view 
 
 \State \textbf{If} there exists greatest $v\geq \mathtt{view}_i$ s.t.\  $M_i$ contains at least $f+1$ end-view $v$ messages \textbf{then}:  \label{15} 
 
 \State \hspace{0.3cm} Form a $(v+1)$-certificate and send it to all processes; 
 
 \State  \textbf{If} there exists some greatest $v>\mathtt{view}_i$ such that either: 
 
  \State \hspace{0.3cm} (i) $M_i$ contains a $v$-certificate $q$, or (ii) $Q_i$ contains $q$ with $q.\text{view}=v$, \textbf{then}: 
 
%  \State \hspace{0.3cm} (ii) $Q_i$ contains $q$ with $q.\text{view}=v$, \textbf{then}: 
  
  \State  \hspace{0.6cm}  Set $\mathtt{view}_i:=v$; Send (either) $q$ to all processes; 
  
  \State  \hspace{0.6cm}  Send all tips $q'$ of $Q_i$ such that $q'.\text{auth}=p_i$ to $\mathtt{lead}(v)$;
  
  \State \hspace{0.6cm}  Send  $(v,q')$ signed by $p_i$  to $\mathtt{lead}(v)$, where $q'$ is a maximal amongst 1-QCs seen by $p_i$ \label{22} 
  
%  \State 
      
       \hrule 
 
 \State \Comment Send 0-votes and 0-QCs 
 
 \State  \textbf{If} $M_i$ contains some $b$ s.t.\  $\mathtt{voted}_i(0,b.\text{type},b.\text{slot},b.\text{auth})=0$:  \label{24} 
 
  \State \hspace{0.3cm} Send a 0-vote for $b$ (signed by $p_i$) to $b.\text{auth}$;  Set  $\mathtt{voted}_i(0,b.\text{type},b.\text{slot},b.\text{auth}):=1$; 
  
   \State  \textbf{If} $M_i$ contains a 0-quorum for some $b$ s.t.:
   
     \State \hspace{0.3cm} (i) $b.\text{auth}=p_i$, \textbf{and} (ii) $p_i$ has not previously sent a 0-QC for $b$ to other processors, \textbf{then}:
 
 % \State \hspace{0.3cm} (ii) There is no $q\in Q_i$ with $q.\text{b}=b$, \textbf{then}:
  
    \State  \hspace{0.6cm}  Send a 0-QC for $b$ to all processes; \label{28} 
    
  %    \State 
      
       \hrule 
 
 \State \Comment Send out a new transaction block
 
 \State \textbf{If} $\mathtt{PayloadReady}_i=1$ \textbf{then}:  \label{30} 
 
   \State \hspace{0.3cm}   $\mathtt{MakeTrBlock}_i$; \label{31} 
   
   %      \State 
      
       \hrule 
 
 \State \Comment Send out a new leader block
 
 \State \textbf{If} $p_i=\mathtt{lead}(\mathtt{view}_i)$, $\mathtt{LeaderReady}_i=1$, $\mathtt{phase}_i(\mathtt{view}_i)=0$ \textbf{and} $Q_i$ does not have a single tip: \label{33} 
 
   \State \hspace{0.3cm}   $\mathtt{MakeLeaderBlock}_i$; \label{34}

 %          \State 
      
       \hrule 
 
 \State \Comment Send 1 and 2-votes for transaction blocks

  \State \textbf{If} there exists $b\in M_i$ with $b.\text{type}=\text{lead}$ and $b.\text{view} =\mathtt{view}_i$ \textbf{and}  \label{36} 
  
 \State there does not exist unfinalized $b\in M_i$ with $b.\text{type}=\text{lead}$ and $b.\text{view} =\mathtt{view}_i$ \textbf{then}:  \label{37} 
 
       \State \hspace{0.3cm} \textbf{If} there exists $b\in M_i$  with $b.\text{type}=\text{Tr}$, $b.\text{view} =\mathtt{view}_i$ and which is a single tip of $M_i$ s.t.:
  
  \State  \hspace{0.6cm}   (i) $b.\text{1-QC}$ is greater than or equal to every 1-QC in $Q_i$ and;
  
  \State  \hspace{0.6cm}    (ii) $\mathtt{voted}_i(1,\text{Tr}, b.\text{slot},b.\text{auth})=0$, \textbf{then}: 
  
    \State \hspace{0.9cm} Send a 1-vote for $b$ to all processes; Set  $\mathtt{phase}_i(\mathtt{view}_i):=1$;
    
        \State \hspace{0.9cm} Set  $\mathtt{voted}_i(1,\text{Tr}, b.\text{slot},b.\text{auth}):=1$;

  \State \hspace{0.3cm} \textbf{If} there exists a 1-QC $q\in Q_i$ which is a single tip of $Q_i$ s.t.:
  
  \State  \hspace{0.6cm}   (i) $q.\text{type}=\text{Tr}$ \textbf{and}   (ii) $\mathtt{voted}_i(2,\text{Tr}, q.\text{slot},q.\text{auth})=0$, \textbf{then}:

  \State  \hspace{0.9cm}  \textbf{If} there does not exist $b\in M_i$ of height greater than $q.\text{h}$:

%  \State  \hspace{0.6cm}    (ii) $\mathtt{voted}_i(2,\text{Tr}, q.\text{slot},q.\text{auth})=0$, \textbf{then}: 
  
    \State \hspace{1.2cm} Send a 2-vote for $q.\text{b}$ to all processes; Set  $\mathtt{phase}_i(\mathtt{view}_i):=1$;
    
        \State \hspace{1.2cm} Set  $\mathtt{voted}_i(2,\text{Tr}, q.\text{slot},q.\text{auth}):=1$; \label{47} 
        
      %       \State \hspace{0.9cm} \textbf{If} $\mathtt{phase}_i(\mathtt{view}_i)=0$, set  $\mathtt{phase}_i(\mathtt{view}_i):=1$;

               %   \State \hspace{0.9cm} \textbf{If} $\mathtt{phase}_i(\mathtt{view}_i)=0$, set  $\mathtt{phase}_i(\mathtt{view}_i):=1$;
               
    %                      \State 
      
       \hrule 
 
 \State  \Comment Vote for a leader block
 
  \State \textbf{If} $\mathtt{phase}(\mathtt{view}_i)=0$:  \label{49} 
 
 \State \hspace{0.3cm}  \textbf{If} $\exists b\in M_i$ with $b.\text{type}=\text{lead}$, $b.\text{view} =\mathtt{view}_i$,  $\mathtt{voted}_i(1,\text{lead}, b.\text{slot},b.\text{auth})=0$ \textbf{then}: 
 
     \State \hspace{0.6cm} Send a 1-vote for $b$ to all processes;  Set  $\mathtt{voted}_i(1,\text{lead}, b.\text{slot},b.\text{auth}):=1$;
     
      \State \hspace{0.3cm}   \textbf{If} $\exists q\in Q_i$ which is a 1-QC with $\mathtt{voted}_i(2,\text{lead}, q.\text{slot},q.\text{auth})=0$, $q.\text{type}=\text{lead}$, 
      \State  \hspace{0.3cm}   $q.\text{view} =\mathtt{view}_i$,  \textbf{then}: 
 
     \State \hspace{0.6cm} Send a 2-vote for $q.\text{b}$ to all processes;  Set  $\mathtt{voted}_i(2,\text{lead}, q.\text{slot},q.\text{auth}):=1$; \label{54} 
     
        %                       \State 
      
       \hrule 
 
 \State \Comment Complain 
 
   \State \textbf{If} $\exists q\in Q_i$ which is maximal according to $\succeq$ amongst those that have not been finalized for time $6\Delta$ since entering view $\mathtt{view}_i$:  \label{56} 
   
      \State \hspace{0.3cm} Send $q$ to $\mathtt{lead}(\mathtt{view}_i)$ if not previously sent; \label{57} 
      
         \State \textbf{If} $\exists q\in Q_i$ which has not been finalized for time $12\Delta$ since entering view $\mathtt{view}_i$: \label{comp}

         \State \hspace{0.3cm}  Send  the end-view message $(\mathtt{view}_i)$ signed by $p_i$ to all processes;  \label{59}

\end{algorithmic}
\end{algorithm}

\subsection{Pseudocode walk-through} \label{wt} 

\textbf{Lines \ref{15}-\ref{22}}: These lines are responsible for view changes. If $p_i$ has received a quorum of end-view $v$ messages for some greatest  $v$ greater than or equal to its present view, then it will use those to form a $(v+1)$-certificate and will send that certificate to all processes (immediately regarding that certificate as received and belonging to $M_i$). 
Upon seeing that it has received a $v$-certificate for some greatest view $v$ greater than its present view, $p_i$ will: (i)  enter view $v$, (ii) send that $v$-certificate to all processes, and (iii) send a view $v$ message to the leader of view $v$, along with any tips of $Q_i$ corresponding to its own blocks. Process $p_i$  will also do the same upon seeing $q$ with $q.\text{view}$ greater than its present view: the latter action ensures that any block $b$ produced by $p_i$ during view $v$ does not point to any $b'$  with $b'.\text{view}>b.\text{view}$. 

\vspace{0.2cm} 
\noindent \textbf{Lines \ref{24}-\ref{28}}. These lines are responsible for the production of 0-QCs. Upon producing any block, $p_i$ sends it to all processes. Providing $p_i$ is correct, meaning that the block is correctly formed etc, other processes will then  send back a 0-vote for the block to $p_i$, who will form a $0$-QC and send it to all processes. 

\vspace{0.2cm} 
\noindent \textbf{Lines \ref{30} and \ref{31}}. These lines are responsible for producing new transaction blocks. Line 30 checks to see whether $p_i$ is ready to produce a new transaction block, before line 31 produces the new block: $\mathtt{PayloadReady}_i$ and $\mathtt{MakeTrBlock}_i$ are specified in Section \ref{formalspec}.

\vspace{0.2cm} 
\noindent \textbf{Lines \ref{33} and \ref{34}}. These lines are responsible for producing new leader blocks. Line \ref{33} ensures that only the leader is asked to produce leader blocks, that it will only do so once ready (having received QCs for previous leader blocks, as required), and only when required to (only if $Q_i$ does not have a single tip and if still  in the first phase of the view). $\mathtt{LeaderReady}_i$ and $\mathtt{MakeLeaderBlock}_i$ are specified in Section \ref{formalspec}. 

\vspace{0.2cm} 
\noindent \textbf{Lines \ref{36}-\ref{47}}. These lines are responsible for determining when correct processes produce 1 and 2-votes for transaction blocks. Lines \ref{36} and \ref{37} dictate that no correct process produces 1 or 2-votes for transaction blocks while in view $v$ until at least one leader block for the view has been finalized (according to the messages they have received), and only if there do not exist unfinalized leader blocks for the view. Given these conditions, $p_i$ will produce a 1-vote for any transaction block $b$ that is a single tip of $M_i$, so long as $b.\text{1-QC}$ is greater than or equal to any 1-QC it has seen. It will produce a 2-vote for a transaction block $b$ if there exists $q$ with $q.\text{b}=b$ which is a  single tip of $Q_i$ and if $p_i$ has not seen any block of greater height. The latter condition is required to ensure that $p_i$ cannot produce a 1-vote for some $b'$ of greater height than $b$, and then produce a 2-vote for $b$ (this fact is used in the proof of Theorem \ref{cons}).  After producing any 1 or 2-vote for a transaction block while in view $v$, $p_i$ enters the second phase of the view and will no longer produce 1 or 2-votes for leader blocks while in view $v$. 

\vspace{0.2cm} 
\noindent \textbf{Lines \ref{49}-\ref{54}}.  These lines are responsible for determining when correct processes produce 1 and 2-votes for leader blocks. Correct processes will only produce such votes while in the first phase of the view. 

\vspace{0.2cm} 
\noindent \textbf{Lines \ref{56}-\ref{59}}. These lines are responsible for the production of new-view messages. The proof of Theorem \ref{live} justifies the choice of $6\Delta$ and $12\Delta$. 
%The choice of $7\Delta$ is justified as follows. When $p_i$ sends $q$ to the leader at time $t$, it is received by time $t+\Delta$. If $q$ has not been finalized yet, and if the leader is correct, then they will produce a leader block observing $q.\text{b}$ by $t+3\Delta$, which will be finalized by time $t+6\Delta$. 

 \section{Establishing consistency and liveness}  \label{cl} 
 
Let $M^*$ be the set of all messages received by any process during the execution. Towards establishing consistency, we first prove the following lemma. 
 
 \begin{lemma} \label{6.1}
 If $q,q'\in M^*$ are 1-QCs with $q\leq q'$ and $q'\leq q$, then $q.\text{b}=q'.\text{b}$. 
 \end{lemma} 
 \begin{proof} 
 Suppose $q.\text{view}=q'.\text{view}$,  $q.\text{type}=q'.\text{type}$, and  $q.\text{h}=q'.\text{h}$. Consider first the case that $q.\text{b} $ and $q'.\text{b} $ are both leader blocks for the same view. 
If $q.\text{slot}=q'.\text{slot}$, but $q.\text{b}\neq q'.\text{b}$, then no correct process can produce 1-votes for both blocks. This gives an immediate contradiction, since two subsets of $\Pi$ of size $n-f$ must have a correct process in the intersection, meaning that 1-QCs cannot be produced for both blocks. So, suppose that $q'.\text{slot}>q.\text{slot}$. Since each leader block $b$ with $b.\text{slot}=s>0$ must point to a leader block $b'$ with $b'.\text{auth}=b.\text{auth}$ and $b'.\text{slot}=s-1$, it follows that $q'.\text{h}>q.\text{h}$, which also gives a contradiction. 
 
 So, consider next the case that $q.\text{b} $ and $q'.\text{b} $ are distinct transaction blocks. Since both blocks are of the same height, and since any correct process only votes for a block when it is a sole tip of its local value $M_i$, no correct process can vote for both blocks. Once again, this gives the required contradiction. 
 \end{proof} 
 
 Note that Lemma \ref{6.1} also suffices to establish a similar result for 2-QCs, since no block can receive a 2-QC without first receiving a 1-QC: No correct process produces a 2-vote for any block without first receiving  a 1-QC for the block.  
 
 \vspace{0.2cm} 
 Lemma \ref{6.1} suffices to show that we can think of all 1-QCs $q\in M^*$ as belonging to a hierarchy, ordered by $q.\text{view}$, then by $q.\text{type}$, and then by $q.\text{h}$, such that if $q$ and $q'$ belong to the same level of this hierarchy then $q.\text{b}=q'.\text{b}$. 
 
 \begin{theorem} \label{cons} 
 The Morpheus protocol satisfies consistency. 
 \end{theorem} 
 \begin{proof} 
  Given the definition of $\mathcal{F}$ from Section \ref{intuition}, let us say $b'\rightarrow b$ iff: 
 \begin{itemize} 
 \item $b'=b$, or;
 \item $b'\neq b_g$ and $b''\rightarrow b$, where $q=b'.\text{1-QC}$ and $b''=q.\text{b}$.
 \end{itemize} 
 
 \noindent To establish consistency it suffices to show the following: 

\begin{enumerate} 
\item[$(\dagger)$:]  If $b$ has a 1-QC  $q_1\in M^*$ and also a 2-QC $q_2\in M^*$, then for any 1-QC  $q\in M^*$ such that $q\geq q_1$, $q.\text{b}\rightarrow b$. 
\end{enumerate} 

Given $(\dagger)$, suppose $M_1\subseteq M_2 \subseteq M^*$. For each $i\in \{ 1,2\}$, let $M_i'$ be the largest set of blocks in $M_i$ that is downward closed (in the sense specified in Section \ref{intuition}). Let $q'_i$ be a maximal  2-QC  in $M_i$ such that $q_i'.\text{b}\in M_i'$, and set $b_i^*=q_i'.\text{b}$, or if there is no such 2-QC  in $M_i$, set $b_i^*=b_g$. 
Let  the sequence $b_k,\dots, b_1=b_g$ be such that $b_k=b^*_2$, and, for each $ j<k$, if $q=b_{j+1}.\text{1-QC}$, then $q.\text{b}=b_{j}$. From $(\dagger)$ it follows that $b^*_1$ belongs to the sequence $b_k,\dots, b_1$, so that $\mathcal{F}(M_2) \supseteq \mathcal{F}(M_1)$. 

\vspace{0.2cm} 
We establish $(\dagger)$ by induction on the level of the hierarchy to which $q$ belongs. If $q\leq q_1$ (and $q_1\leq q$) then the result follows from Lemma \ref{6.1}. 

\vspace{0.2cm} 
For the induction step, suppose that $q>q_1$ and suppose first that $q.\text{type}=\text{lead}$. Let $s=q.\text{slot}$, $v=q.\text{view}$. By validity of $q.\text{b}$, if $s>0$, $q.\text{b}$ points to a unique $b^*$ with $b^*.\text{type}=\text{lead}$,  $b^*.\text{auth}=q.\text{auth}$ and $b^*.\text{slot}=s-1$.  If $s=0$ or $b^*.\text{view}<v$, then $q.\text{just}$ (i.e. $(q.\text{b}).\text{just}$) contains $n-f$ view $v$ messages, each signed by a different process in $\Pi$. Note that, in this case,  any correct process that produces a 2-vote for $b$ must do so before sending a view $v$ message. It follows that, in this case,  $q.\text{1-QC}$ (i.e. $(q.\text{b}).\text{1-QC}$) belongs to a level of the hierarchy strictly below $q$ and greater than or equal to that of $q_1$. The result therefore follows by the induction hypothesis.  If $s>0$ and $b^*.\text{view}=v$, then $q.\text{1-QC}$ is a 1-QC-certificate for $b^*$. Once again, $q.\text{1-QC}$ therefore belongs to a level of the hierarchy strictly below $q$ and greater than or equal to that of $q_1$, so that the result follows by the induction hypothesis. 

So,  suppose next that $q.\text{type}=\text{Tr}$.  Note that, in this case,  any correct process that produces a 2-vote for $b$ must do so before sending a 1-vote for $q.\text{b}$. If $q.\text{view}>b.\text{view}$  this follows immediately, because a correct process $p_i$ only sends 1 or 2-votes for any block $b'$ while  $\mathtt{view}_i=b'.\text{view}$. If  $q.\text{view}=b.\text{view}$ and $b.\text{type}=\text{lead}$, this follows because no correct process sends 1 or 2-votes for a leader block after having voted for a transaction block within the same view.  If  $q.\text{view}=b.\text{view}$ and $b.\text{type}=\text{Tr}$, this follows because any correct process only sends a 2-vote for $b$ so long as there does not exist $b'\in M_i$ of height greater than $b$. Also, any
correct process that produces a 2-vote for $b$ will not vote for $q.\text{b}$ unless $q.\text{1-QC}$ is greater than or equal to any 1-QC it has received. It follows that $q.\text{1-QC}$ belongs to a  level of the hierarchy strictly below $q$ and greater than or equal to that of $q_1$. Once again,  the result follows by the induction hypothesis. 
 \end{proof} 
 
 \begin{theorem} \label{live} 
 The Morpheus protocol satisfies liveness. 
 \end{theorem} 
 \begin{proof} 
 Towards a contradiction, suppose that correct $p_i$ produces a transaction block $b$, which never becomes finalized (according to the messages received by $p_i$). Note that all correct processes eventually send 0-votes for $b$ to $p_i$, meaning that $p_i$ forms a 0-certificate for $b$, which is eventually received by all correct processes. 
 Since correct processes send end-view messages if a some QC is not finalized for sufficiently long within any given view (see line \ref{comp}), correct processes must therefore enter infinitely many views. Let $v$ be a view with a correct leader, such that the first correct process $p_j$ to enter view $v$ does so at some timeslot  after GST, and after $p_i$ produces $b$. Process $p_j$ sends a v-certificate to all processes upon entering the view, meaning that all correct processes enter the view within time $\Delta$ of $p_j$ doing so. Upon entering view $v$, at time $t$ say, note that $p_i$ will send a QC for a transaction block $b'$ that it has produced to the leader. This block $b'$  has a slot number greater than or equal to that of $b$. The leader will produce a leader block observing $b'$ by time $t+3\Delta$, which will be finalized (according to the messages received by $p_i$) by time $t+6\Delta$.
    \end{proof} 
    
\section{Latency and complexity analysis}     \label{metrics}

In this section, we discuss latency and communication complexity for Morpheus. As is standard for protocols that operate in partial synchrony, we focus on values for these metrics after GST. In analysing latency for DAG-based protocols, some papers focus on the number of `layers' of the DAG of blocks required for finalization. This hides information, such as the amount of time required to form each layers. We therefore consider the time to finalization expressed in terms of $\Delta$  and $\delta$ (recall that $\delta$ is the unknown and actual least upper-bound on message delay after GST). 

\vspace{0.2cm} 
\noindent \textbf{Worst-case latency}. If $p_i$ produces a block before entering a view $v$ which begins after GST, and if the block is not finalized while $p_i$ is still in view $v$, then that view must have a faulty leader (if the leader of the view is correct, then the block will be finalized in time $O(\delta)$ of $p_i$ entering the view and while $p_i$ is still in that view). Any such view is of length $O(\Delta)$. Worst-case latency is therefore $O(f_a\Delta)$, where $f_a$ is the actual (unknown) number of faulty processes. 

\vspace{0.2cm} 
\noindent \textbf{Worst-case latency during high throughput in a view with a correct leader}. If correct $p_i$ produces a transaction block $b$ at timeslot $t$, then $p_i$ will produce a 0-QC for $b$ by $t+2\delta$, which will be received by the leader of the view by $t+3\delta$. The leader may then produce a new block immediately, but, in the worst case, at most $2\delta$ will then pass before the leader produces the next leader block for the view. That leader block (and $b$) will then be finalized (according to the messages received by $p_i$) by time $t+ 8\delta$. 

\vspace{0.1cm} 
\noindent \emph{Optimisations}. The analysis above considers the \emph{worst-case} latency during high throughput while in a view with a correct leader. As described in the Autobahn paper, however, there are a number of optimisations (with corresponding trade-offs) that can be implemented so as to reduce latency in certain `good cases'. Note first that we can reduce the bound of $t+ 8\delta$  to $t+7\delta$ by having processes send 0-votes for $b$ `to all', and then having each process form their own 0-QC for $b$.  
(Whether doing so would reduce \emph{real} latency depends on subtle issues such as connection speeds, and so on.) Autobahn also specifies (see Section 5.52 of \cite{giridharan2024autobahn}) an optimisation whereby a leader may point to a block $b$ (just with a hash rather than a QC) immediately upon receiving it if they believe the block producer to be trustworthy. Other processes will then only vote for the leader block upon receiving a $0$-QC for $b$. Thus reduces the bound to $t+6\delta$ in the `good case' that all blocks pointed to do receive 0-QCs. Another standard optimisation \cite{gueta2019sbft,kotla2007zyzzyva} (also discussed in the `Fast Path' paragraph of Section 5.2.1 in \cite{giridharan2024autobahn}) allows a leader block to be finalised after receiving a 1-QC formed from $n$ votes, which reduces the bound to $t+5\delta$ in the case that all processes are acting correctly while in the view. Finally, we assumed in the analysis above that $2\delta$ elapses from the time that the leader receives a new transaction block $b$ until the leader produces a new leader block. In the good case that the leader produces the new leader block immediately, the previous bound of $t+5\delta$ reduces to $t+3\delta$.

\vspace{0.2cm} 
\noindent \textbf{Worst-case latency during low throughput}. We note that a Byzantine adversary can issue transactions causing high throughput. In analysing latency during low throughput, it therefore really makes sense to consider benign faults (omission/crash failures). 

If $p_i$ issues a transaction block $b$ at time $t$, and if no transaction blocks conflict with $b$, then $b$ will be finalized (according to the messages received by $p$) by $t+3\delta$. We note that, in a context where each process produces their own transactions, this actually reduces latency by at least $\delta$ when compared to protocols such as PBFT and Tendermint, since there is no need to send transactions to the leader before they can be included in a block.

\vspace{0.2cm} 
\noindent \textbf{Amortised complexity during high throughput}. It is common \cite{keidar2021all,miller2016honey} to show that one can achieve linear amortised communication complexity (i.e.\ communication complexity per transaction) for DAG-based protocols by using \emph{batching}, i.e.\ by requiring that each block contain a large number of transactions, which may be $O(n)$ or greater. The downside of this approach is that, depending on the scenario, waiting for a set of transactions of sufficient size before issuing a block may actually increase latency. The use of erasure codes \cite{alhaddad2022balanced,nayak2020improved} is also sometimes required, which introduces further subtle trade-offs: these cryptographic methods introduce their own latencies. Like Autobahn, Morpheus achieves linear amortised communication complexity during high throughput, without the need for batching or the use of erasure coding. 

For the sake of concreteness, suppose that each correct process produces one transaction block in each interval of length $\delta$, and suppose $\Delta$ is $O(\delta)$. Suppose correct $p_i$ issues a transaction block $b$ at time $t_1$ while in view $v_1$, and that the first correct process to enter $v_1$ does so after GST.  Suppose $b$  is ultimately finalized at time $t_2$ while $p_i$ is in view $v_2\geq v_1$, meaning that all views in $ [ v_1,v_2)$ have Byzantine leaders. Let $k=v_2-v_1+1$. We consider the total communication cost for correct processes between $t_1$ and $t_2$, and amortise by dividing that cost  by the total number of new transactions finalized during this interval. For any transaction block $b'$ issued  during this interval, sending $b'$ to all processes and forming a 0-QC for $b'$ induces linear cost per transaction. For each view in $ [ v_1,v_2)$, the communication cost induced (between $t_1$ and $t_2$) by the sending of 1 and 2-votes by correct processes is $O(n^2)$. So, overall, such messages contribute $O(kn^2)$ to the communication cost. The messages sent in lines \ref{56}-\ref{59} of the pseudocode similarly contribute an overall $O(kn^2)$ to the communication cost, as do view certificates, new-view messages  and all messages sent in lines \ref{15}-\ref{22} of the pseudocode. Leader blocks sent by the leader of view $v_2$ prior to $t_2$ induce $O(n^2)$ communication cost. Since $\Omega(kn)$ transactions are finalized in the interval $[t_1,t_2]$, this gives an overall amortised communication cost that is $O(n)$.  

\vspace{0.2cm} 
\noindent \textbf{Amortised complexity during low throughput}. In low throughput, sending a block of constant size to all processes gives a communication cost $O(n)$, while the sending of 1 and 2-votes for the block gives cost $O(n^2)$ (since votes are sent `to all'). This gives an amortised cost which is $O(n^2)$ per transaction. An $O(n)$ bound can be achieved, either by using batching, or else by having votes be sent only to the block producer, rather than to all processes, and then requiring the block producer to distribute 1 and 2-QCs (increasing latency by $2\delta$). We do not use the latter option, since doing so seems like  a false economy: the leader is anyway the `bottleneck' in this context, which means that in real terms using this option is only likely to increase latency. 

\vspace{0.2cm} 
\noindent \textbf{Latency comparison with Autobahn and Sailfish}. Morpheus is essentially the same as Autobahn during high throughput, and so has all the same advantages as Autobahn (low latency and \emph{seamless} recovery from periods of asynchrony) in that context, while also being quiescent and giving much lower latency during low throughput. `Seamless recovery' is a concept discussed in the Autobahn paper, and we do not repeat those discussion here. 

We use Sailfish as a representative when making comparisons with DAG-based protocols, since (as far as we are aware) Sailfish has the lowest latency amongst such protocols at the present time, at least in the `good case' that leaders are honest. In fact, apples-to-apples comparisons with Sailfish are difficult, because there are many cases to consider when analysing Sailfish that do not have analogues in the context of analysing Morpheus. Let us suppose, for example, that `leaders' are correct. Then, for Sailfish,   the time it takes for a block to be finalized will depend on whether it happens to be one of the $n-f$ blocks pointed to by the next leader block. In Sailfish, each block is initially reliably broadcast, but there are also a number of ways in which reliable broadcast can be implemented. 
In Morpheus, we have required that $0$-votes be sent only to the block producer (so as to limit communication complexity), but one could alternatively have $0$-votes be sent to all processes, reducing worst-case latency by $\delta$, and similar considerations apply when implementing reliable broadcast. To make a comparison that is as fair as possible, we therefore assume that all-to-all communication is used when carrying out reliable broadcast (as in Sailfish), meaning that it takes time at least $2\delta$ when the broadcaster is correct. Correspondingly, we then make overall comparisons with the total worst-case bound of $7\delta$ for Morpheus given in the paragraph on `optimisations' above, for the case of a view with a correct leader (we do not consider the further optimisations for Morpheus that it was previously noted can bring the latency down to $3\delta$ in the `good case'). We are also generous to Sailfish in considering the latency for blocks that are amongst the $n-f$ blocks pointed to by the leader of the next round. With these assumptions, the latency analyses are then essentially identical for the two protocols. For Sailfish, reliably broadcasting a block takes time $2\delta$. Upon receiving that block, the next leader may take time up to $2\delta$ to produce their next block, which will then be finalized within a further time $3\delta$. This gives the same worst-case  latency bound of $7\delta$ when leaders are correct. Advantages of Morpheus over Sailfish include the fact that it has linear amortised communication complexity without batching, and much lower latency of $3\delta$ during low throughput.

\section{Related Work} \label{rw} 
 % \cite{castro1999practical} and Tendermint \cite{buchman2016tendermint,buchman2018latest},
 % To mention: PBFT,  Zyzzyva, Tendermint, Hotstuff, Hashgraph, Aleph, Narwhal, DAG-Rider, Tusk, Bullshark, Cordial Miners, Mysticeti, BBCA-chain, Shaol, Shaol++ (with discussion of relationship to Sailfish), Sailfish, Motorway, Autobahn (and some differences with Motorway), GradedDAG [17] and LightDAG [15] (which use consistent broadcast rather than RB), Avalanche
 % Reliable broadcast (two references) 

 Morpheus uses a PBFT \cite{castro1999practical} style approach to view changes, while consistency between finalised transaction blocks within the same view uses an approach similar to Tendermint \cite{buchman2016tendermint,buchman2018latest} and Hotstuff \cite{yin2019hotstuff}. As noted in Section \ref{metrics}, Hotstuff's approach of relaying all messages via the leader could be used by Morpheus during low throughput to decrease communication complexity, but this is unlikely to lead to a decrease in `real' latency (i.e.\ actual finalisation times). As also noted in Section \ref{metrics}, the optimistic `fast commit' of Zyzzyva \cite{gueta2019sbft,kotla2007zyzzyva} can also be applied as a further optimisation. 

 \vspace{0.2cm} 
 Morpheus transitions between being a leaderless `linear' blockchain during low throughput to a leader-based DAG-protocol during high throughput. DAG protocols have been studied for a number of years, Hashgraph \cite{baird2016swirlds} being an early example. Hashgraph builds an unstructured DAG and suffers from latency exponential in the number of processes. Spectre was another early DAG protocol, designed for the `permissionless'' setting \cite{sompolinsky2016spectre}, with proof-of-work as the mechanism for sybil resistance. The protocol implements a `payment system', but does not totally order transactions.  Aleph \cite{gkagol2019aleph} is more similar to most recent DAG protocols in that it builds a structured DAG in which each process proceeds to the next `round' after receiving blocks from $2f+1$ processes corresponding to the previous round, but still has greater  latency than modern DAG protocols.  

 \vspace{0.2cm}
 More recent DAG protocols use a variety of approaches to consensus. Narwhal \cite{danezis2022narwhal} builds a DAG for the purpose of ensuring data availability, from which (one option is that) a protocol like Hotstuff or PBFT can then be used to efficiently establish a total ordering on transactions. DAG-Rider \cite{keidar2021all}, on the other hand, builds the DAG in such a way that a total ordering can be extracted from the structure of the DAG, with zero further communication cost. The protocol proceeds in `waves', where each wave consists of four rounds, each round building one `layer' of the DAG. In each round, each process uses an instance of Reliable Broadcast (RBC) to disseminate their block for the round. Each wave has a leader and an 
 expected six rounds (6 sequential RBCs) are required to finalise
the leader's block for the first round of the wave. This finalises all blocks observed by that leader block, but other blocks (such as those in the same round as the leader block) may have signicantly greater latency. Tusk \cite{danezis2022narwhal} is an implementation based on DAG-Rider. 

 \vspace{0.2cm} 
 Given the ability of DAG-Rider to handle significantly higher throughput in many settings, when compared to protocols like PBFT that build a linear blockchain, much subsequent work has taken a similar approach, while looking to improve on latency. While DAG-Rider functions in asynchrony, Bullshark \cite{spiegelman2022bullshark} is designed to achieve lower latency in the partially synchronous setting. 
 GradedDAG \cite{dai2023gradeddag} and LightDAG \cite{dai2024lightdag} function in asynchrony, but look to improve latency by replacing RBC \cite{bracha1987asynchronous} with weaker primitives, such as consistent broadcast \cite{srikanth1987simulating}. This means that those protocols solve Extractable SMR (as defined in Section \ref{esmr}), rather than SMR, and that further communication may be required to ensure full block dissemination in executions with faulty processes. Cordial Miners \cite{keidar2022cordial} has versions for both partial synchrony and asynchrony and further decreases latency 
by using the DAG structure (rather than any primitive such as Consistent or Reliable Broadcast) for equivocation exclusion. Mysticeti \cite{babel2023mysticeti} builds on Cordial Miners and establishes a mechanism to  accommodate multiple leaders within a single round. Shoal \cite{spiegelman2023shoal} and Shoal++ \cite{arun2024shoal++} extend Bullshark by establishing a `pipelining approach' that implements simultaneous instances of Bullshark with a leader in each round. This reduces latency in the good case because one is required to wait less time before reaching a round in which a leader block is finalised. Both of these papers, however, use a `reputation' system to select leaders, which comes with its own trade-offs. Sailfish \cite{shrestha2024sailfish} similarly describes a mechanism where each round has a leader, but does not make use of a reputation system. As noted previously, the protocol most similar to Morpheus during high throughput is Autobahn \cite{giridharan2024autobahn}. One of the major distinctions between Autobahn and those previously discussed, is that most blocks are only required to point to a single parent. This significantly decreases communication complexity when the number of processes is large and allows one to achieve linear ammortised communction complexity without the use of erasure coding \cite{alhaddad2022balanced,nayak2020improved} or batching \cite{miller2016honey}.

 \bibliographystyle{plainurl}

%
%      \vspace{0.1cm} 
% \noindent \textbf{Signatures}. We write $m_{p_i}$ to denote\footnote{While the subfix $p_i$ is sometimes used to indicate a value specific to $p_i$ (such as in the case of $\mathtt{log}_{p_i}$), no ambiguity will result. The subfix $p_i$ only indicates a signature when applied to individual messages.} the message $m$ signed by $p_i$. 
%  

\end{document}